\documentclass{cta-author}
\usepackage{amsmath}
\usepackage{verbatim}
\usepackage{ragged2e}
\usepackage{pifont}
\usepackage{subcaption}
\usepackage{caption,psfrag}
\usepackage{amssymb}
\usepackage{amsthm}
\usepackage{float}
\usepackage{mathtools}
\usepackage{mathtools}
\usepackage[english]{babel}
\newtheorem{theorem}{Theorem}[section]

\newtheorem{lemma}[theorem]{Lemma}

\usepackage{amsthm}
\theoremstyle{plain}
\newtheorem{assumption}{Assumption}
{}
\newtheorem{prop}{Proposition}

\begin{document}


\title{$\text{H}_{\infty}$ Tracking Control via Variable Gain Gradient Descent-Based Integral Reinforcement Learning for Unknown Continuous Time Nonlinear System}

\author{\au{Amardeep Mishra$^{1}$}, \au{Satadal Ghosh$^{2\corr}$}}

\address{\add{1}{Student, Department of Aerospace Engineering, IIT Madras, Chennai 600036, India}
\add{2}{Faculty, Department of Aerospace Engineering, IIT Madras, Chennai 600036, India}
\email{satadal@iitm.ac.in}}



\begin{abstract}
Optimal tracking of continuous time nonlinear systems has been extensively studied in literature. However, in several applications, absence of knowledge about system dynamics poses a severe challenge to solving the optimal tracking problem. 
This has found growing attention among researchers recently, and integral reinforcement learning (IRL)-based method augmented with actor neural network (NN) have been deployed to this end. However, very few studies have been directed to model-free $H_{\infty}$ optimal tracking control that helps in attenuating the effect of disturbances on the system performance without any prior knowledge about system dynamics. To this end a recursive least square-based parameter update was recently proposed. However, gradient descent-based parameter update scheme is more sensitive to real-time variation in plant dynamics. And experience replay (ER) technique has been shown to improve the convergence of NN weights by utilizing past observations iteratively. Motivated by these, this paper presents a novel parameter update law based on variable gain gradient descent and experience replay technique for tuning the weights of critic, actor and disturbance NNs. 
The presented update law leads to improved model-free tracking performance under $\mathcal{L}_2$-bounded disturbance. Simulation results are presented to validate the presented update law.
\end{abstract}

\maketitle

\section{Introduction}\label{sec1}
Optimal control is one of the prominent control techniques that aims to find control policies that minimizes a cost function subjected to plant dynamics as constraints. 
Traditional optimal control techniques require full knowledge of plant dynamics and corresponding parameters for their implementation. However, in practice knowledge of the same might be partially available or unavailable. In order to implement optimal control methods online under such limitations, reinforcement learning (RL) \cite{sutton1998introduction}\cite{lewis2013reinforcement} and adaptive dynamic programming (ADP) \cite{powell2007approximate}\cite{zhang2012adaptive} approaches were proposed that solve optimal control problem forward in time. 

Regulation problems and trajectory tracking problems are the two broad classifications of the optimal control problem. The prime objective of regulation problems \cite{murray2002adaptive,abu2005nearly,li2012optimal,yang2014online,zhao2014mec,zhu2015convergence,bhasin2013novel} is to find a control policy that brings the desired states to origin in finite amount of time while minimizing a cost function. On the other hand, optimal tracking control problems (OTCP) \cite{park1996optimal, toussaint2000h, alameda2007optimal, zhang2011data} entails finding control policies that will make the desired states (output of the system) track a time varying reference trajectory. 
Traditionally, the OTCP requires development of two different controllers: (i) transient controller and (ii) steady state control \cite{zhang2012adaptive}, \cite{zhang2011data}. 
Limitation of traditional OTCP solving schemes lies in the requirement of (a) Knowledge of reference dynamics and (b) invertibility condition on control gain matrix. 
Modares et al. \cite{modares2014optimal}, \cite{kiumarsi2014reinforcement} proposed augmented system comprising of error and desired dynamics to by-pass this limitation. Finding the control policy that stabilizes the augmented system while minimizing the performance index was the prime objective of their novel control algorithm. 
The control policy generated by their algorithm also consisted of both transient and steady state controllers.

In the ADP schemes mentioned above identifiers were used to obviate the exact knowledge of nominal plant dynamics. 
However, identifiers add to the computational complexity and also reduce the accuracy of the computations \cite{zhu2016using}. 
In most cases, identifiers also require the knowledge of structure of the plant dynamics. 
Hence, efforts have been devoted to make RL schemes either partially model-free or completely model-free. 
In order to develop continuous time optimal control policies under partial or no knowledge of plant dynamics integral reinforcement learning (IRL) algorithm was leveraged. 
While first few results in this direction for regulation problem were presented in \cite{vrabie2009adaptive,jiang2012computational,luo2013data,vamvoudakis2014online}, Modares et al. \cite{modares2014optimal} developed algorithms for OTCP for partially-unknown system. 
Thereafter, Zhu et al. \cite{zhu2016using} developed Off-policy model-free tracking control of continuous time nonlinear systems using IRL.
They leveraged experience replay (ER) technique to effectively utilize past observations in order to update the NN weights.
Further, the UUB stability of the update law was also proved.
 
It may also be noted that most of the aforementioned RL schemes, for both regulation and tracking problems, do not deal with attenuation of the effects of disturbance. 
To this end, $H_{\infty}$ regulation problem has been studied using RL both offline \cite{abu2008neurodynamic,zhang2011iterative} and online \cite{vamvoudakis2012online,vamvoudakis2011online,modares2014online}. 
Online IRL was also utilized in \cite{vrabie2011adaptive} 
and \cite{luo2014off} for $H_{\infty}$ regulation problem for partially-unknown system.
Note that under partial or no knowledge of the plant dynamics structure, IRL has been leveraged in several literature for regulation problem, while very few studies have dealt with IRL for OTCP problem with disturbance rejection. To the best of the authors' knowledge, \cite{modares2015h} and \cite{zhang2017finite} are the only few papers that have recently presented control policies for model-free OTCP of continuous time nonlinear system in $H_{\infty}$ framework.
Modares et al. \cite{modares2015h} updated the parameters of critic, actor and disturbance NNs using least square method, which could only be initiated after certain number of data had been collected. 
This makes their algorithm less sensitive to real-time variations in plant dynamics \cite{zhu2016using}. 
On the other hand, Zhang et al. \cite{zhang2017finite} utilized gradient descent driven parameter update law for $H_{\infty}$ tracking control problem, and uniform ultimate boundedness (UUB) stability of the parameter update law was proven. 
However, their gradient descent followed a constant learning rate.

While continuous time update law driven by gradient descent is more sensitive to real-time variations in plant dynamics, experience replay (ER) technique has been shown to improve the learning speed significantly by utilizing past observations iteratively \cite{zhu2016using}. 
Also, addition of 'robust terms' in update law was shown to shrink the residual set in \cite{liu2015reinforcement}. 
Inspired by \cite{zhu2016using} and \cite{modares2015h}, this paper presents a novel off-policy IRL-based $H_{\infty}$ tracking control scheme for continuous time nonlinear system, in which the parameter update laws for tuning the weights of critic, actor and disturbance NNs are driven by variable gain gradient descent and ER technique in addition to robust terms. Instead of a constant learning rate in traditional gradient descent-based schemes, the learning rate of gradient descent developed in this paper is variable and a function of Hamilton-Jacobi-Isaac (HJI) error. 
This results in an increased learning rate when HJI error is large and the learning rate is reduced as the HJI error becomes smaller. 
The variable gain gradient descent technique is also shown to have the added advantage of shrinking the size of the residual sets, which the NN weights finally converge to. 
Term corresponding to ER technique and robust term in the update law also contribute in further shrinking the size of the residual set. Unlike \cite{liu2015reinforcement}, the update law presented in this paper, not only leverages robust terms for present instance but also past instances as well.
After the completion of learning phase, the final learnt policies leveraging variable gain gradient descent are executed and they are shown to reduce the oscillations in transient phase and steady-state errors thus resulting in improved tracking performance. 

The rest of the paper is structured as follows. Preliminaries and background of $H_{\infty}$ tracking controller and the tracking HJI equation for augmented system have been presented in Section \ref{hinft}. 
Next, in order to obviate the requirement of system dynamics in policy evaluation step, model-free version of HJI equation to formulate IRL, and neural networks to approximate value function, control and disturbance policies are presented in Section \ref{irl}. 
Section \ref{tuning} highlights the main contribution of this paper, i.e., the continuous time weight update law that is driven by variable gain gradient descent and experience replay technique.
The update law also incorporates the robust terms and their past observations. 
UUB stability analysis for the proposed mechanism is shown. 
Numerical studies are presented in Section \ref{res} to justify the effectiveness of the presented algorithm. 
Finally, Section \ref{conclusion} provides concluding remarks.

\section{$H_{\infty}$ Tracking Problem and HJI Equation}\label{hinft}
\subsection{Problem Formulation}\label{prelim}

It is desired to drive certain states of interest of the dynamical system to follow predefined reference trajectories under $\mathcal{L}_2$-bounded disturbance in an optimal way. Let the dynamical system be described by an affine-in-control differential equation:
\begin{equation}
\begin{split}
\dot{x}=f(x)+g(x)u+k(x)d
\end{split}
\label{dyn}
\end{equation}
where, $x \in \mathbb{R}^n$, $u \in \mathbb{R}^m$, $d \in \mathbb{R}^{q_{1}}$, $f(x):\mathbb{R}^n \to \mathbb{R}^n$ is the drift dynamics, $g(x): \mathbb{R}^n \to \mathbb{R}^{n\times m}$ represents the control coupling dynamics and $k(x):\mathbb{R}^n \to \mathbb{R}^{n\times q_1}$ is disturbance dynamics.
In the subsequent analysis in this paper, it is assumed that none of the system dynamics, that is $f(x),g(x)$ and $k(x)$, are known. However, Lipschitz continuity for the system dynamics as well as controllability of the system over a compact set $\Omega \in \mathbb{R}^n$ are assumed. The reference trajectory is generated by a command generator or a reference system whose dynamics is described by:
\begin{equation}
\dot{x}_d=\eta(x_d)
\end{equation}
Thus, the error is given by:
\begin{equation}
\begin{split}
 e = x-x_d
\end{split}
\end{equation}
Therefore, the error dynamics is given as,
\begin{equation}
\dot{e}=(f(x)+g(x)u+k(x)d-\eta(x_d))
\label{err_dyn}
\end{equation}
In order to formulate corresponding HJI equation and assess the effect of disturbance on the closed-loop system, a virtual performance index ($|X|^2$) is defined as \cite{modares2015h},
\begin{equation}
\begin{split}
|X|^2=e^TQe+u^TRu
\end{split}
\label{fict_cost}
\end{equation}
where, $Q$ and $R$ are positive definite matrices with only diagonal entries. 
It is to be noted that all the vector or matrix norms used in this paper are 2-norm or the Eucledian norm.
In \cite{modares2015h}, the disturbance attenuation condition was characterized as the $\mathcal{L}_2$-gain is smaller than or equal to $\alpha$ for all $d \in L_2[0,\infty)$, that is,
\begin{equation}
\begin{split}
\frac{\int_t^{\infty}e^{-\gamma(\tau-t)}|X|^2d\tau}{\int_t^{\infty}e^{-\gamma(\tau-t)}|d(\tau)|^2d\tau} \leq \alpha^2
\end{split}
\label{l2}
\end{equation}
where $0\leq\gamma$ is the discount factor and $\alpha$ determines the degree of attenuation from disturbance input to the virtual performance measure. The value of $\alpha$ is selected based on trial and error. The minimum value of $\alpha$, for which (\ref{l2}) is satisfied provides optimal-robust control solution \cite{modares2015h}. Now, using (\ref{fict_cost}) and (\ref{l2}),
\begin{equation}
    \begin{split}
 \int_t^{\infty}e^{-\gamma(\tau-t)}(e^TQe+u^TRu)d\tau \leq \alpha^2 \int_t^{\infty}e^{-\gamma(\tau-t)}\|d(\tau)\|^2d\tau      
    \end{split}
    \label{dist_att}
\end{equation}
Finding a control policy $u$ dependent on tracking error and reference trajectory such that the system dynamics (\ref{dyn}) satisfies the disturbance attenuation condition (\ref{dist_att}) and that the error dynamics (\ref{err_dyn}) is locally asymptotically stable for $d=0$ forms $H_{\infty}$ tracking control problem \cite{modares2015h}.

\subsection{HJI Equation: Preliminaries}\label{hji}

The first part of this section deals with the development of HJI equation for solving the  $H_{\infty}$ tracking problem stated above, while the second part discusses about policy iteration steps. As discussed in \cite{modares2015h}, the $H_{\infty}$ tracking problem can also be posed as a min-max optimization problem subjected to augmented system dynamics comprising of error dynamics and desired states dynamics. Subsequently, the solution to min-max optimization problem is obtained by imposing the stationarity condition on the Hamiltonian. In order to formulate tracking HJI equation, an augmented state vector is defined as,
\begin{equation}
z=[e^T,x_d^T]^T
\end{equation}
And, the augmented system dynamics is then given as,
\begin{equation}
\begin{split}
\dot{z}=F(z)+G(z)u+K(z)d
\end{split}
\label{eq:aug}
\end{equation}
Where, 
\begin{equation}
F(z)=\begin{pmatrix}
(f(x)-\eta(x_d)) \\
\eta(x_d)
\end{pmatrix},~G(z)=\begin{pmatrix}
g(x) \\
0
\end{pmatrix},~K(z)=\begin{pmatrix}
k(x) \\
0
\end{pmatrix}
\end{equation}
In the subsequent analysis, $F\triangleq F(z)$, $G\triangleq G(z)$ and $K\triangleq K(z)$.
Using augmented states, the attenuation condition (\ref{dist_att}) can be described as,
\begin{equation}
\int_t^{\infty}e^{-\gamma(\tau-t)}(z^TQ_1z+u^TRu)d\tau \leq \alpha^2\int_t^{\infty}e^{-\gamma(\tau-t)}(\|d(\tau)\|^2)d\tau
\end{equation}
where, $Q_1$ is given by:
\begin{equation}
\begin{split}
Q_1=\begin{pmatrix}
Q_{n\times n} & 0_{n\times n} \\
0_{n\times n} & 0_{n\times n}
\end{pmatrix}_{2n\times 2n}
\end{split}
\end{equation}
Thus, a final performance index consisting of disturbance input is defined as:
\begin{equation}
\begin{split}
J(u,d)&=\int_t^{\infty}e^{-\gamma(\tau-t)}\Big(z(\tau)^TQ_1z(\tau)+u(\tau)^TRu(\tau)\\
&-\alpha^2\|d(\tau)\|^2\Big)d\tau
\end{split}
\label{l2_perf}
\end{equation}
The problem of finding control input $u$ that satisfies (\ref{l2}) is same as minimizing (\ref{l2_perf}) subjected to augmented dynamics. In \cite{bacsar2008h}, a direct relationship between $H_{\infty}$ control problem and two-player zero-sum differential game was established. It was shown that solution of the $H_{\infty}$ control problem is equivalent to solution of the following zero-sum game:
\begin{equation}
\begin{split}
V^*(z)=J(u^*,d^*)=\min_u\max_d J(u,d)
\end{split}
\end{equation}
In the subsequent analysis, $V \triangleq V(z)$, $V_z \triangleq \nabla{V}(z)$. 
The term $J$ is as defined in (\ref{l2_perf}) and $V^*$ is optimal value function.
Existence of game theoretic saddle point was also shown to guarantee the existence of solution of the two-player zero-sum game control problem. This is encapsulated in following Nash condition:
\vspace{-.1cm}
\begin{equation}
\begin{split}
V^*(z)=\min_u\max_d J(u,d)=\max_d\min_u J(u,d)
\end{split}
\end{equation}
\vspace{-.05cm}
Differentiating (\ref{l2_perf}) along augmented system trajectories the following Bellman equation is obtained:
\begin{equation}
\begin{split}
z^TQ_1z+u^TRu-\alpha^2d^Td-\gamma V+V_z(F+Gu+Kd)=0
\end{split}
\label{eq:bell}
\end{equation}
Let the Hamiltonian be defined as:
\begin{equation}
\begin{split}
\mathcal{H}(z,V,u,d)&=z^TQ_1z+u^TRu-\alpha^2d^Td-\gamma V\\
&+V_z(F+Gu+Kd)
\end{split}
\label{hamilt}
\end{equation}
\vspace{-.05cm}
$V^*$ being the optimal cost, satisfies the Bellman equation. Applying stationarity condition on the Hamiltonian, both optimal control input and disturbance input are obtained as follows,
\begin{equation}
\begin{split}
\frac{\partial \mathcal{H}(z,V^*,u,d)}{\partial u}=0 \implies u^*=-\frac{1}{2}R^{-1}G^T\nabla{V^*} \\
\frac{\partial \mathcal{H}(z,V^*,u,d)}{\partial d}=0 \implies d^*=\frac{1}{2\alpha^2}K\nabla{V^*}
\end{split}
\label{opt_ud}
\end{equation}
\vspace{-.06cm}
The optimal control input and disturbance input given above provide saddle point solution to the game \cite{abu2008neurodynamic}. Using (\ref{opt_ud}) in (\ref{hamilt}) tracking HJI equation is,
\begin{equation}
\begin{split}
z^TQ_1z&+V^*_zF-\gamma V^*-\frac{1}{4}V^{*T}_zG^TR^{-1}GV^*_z\\
&+\frac{1}{4\alpha^2}V^{*T}_zKK^TV^*_z=\mathcal{H}(z,V^*,u^*,d^*)=0
\end{split}
\label{eq:tracking_hji}
\end{equation}

\begin{theorem}\label{th:vdot}
Let $V^*$ be the smooth positive semi-definite and quadratic solution of the tracking HJI equation (\ref{eq:tracking_hji}), then the optimal control action ($u^*$) generated by (\ref{opt_ud}) makes the tracking error dynamics (\ref{err_dyn}) asymptotically stable in the limiting sense when discount factor $\gamma \to 0$ and when $d=0$. Additionally, there exists an upper bound for the discount factor $\gamma$ below which the error dynamics is asymptotically stable. 
However, if $\gamma$ is greater than this bound or if $d \neq 0$, then UUB stability can only be ensured.
\end{theorem}
\begin{proof}
The detailed discussion of this proof can be seen in Theorems 3 and 4 of \cite{modares2015h}.
Theorem 3 of \cite{modares2015h} proves the asymptotic stability of the error dynamics in the limiting sense when $\gamma \to 0$. 
Theorem 4 of \cite{modares2015h} on the other hand provides an upper bound over the discount factor $\gamma$ below which asymptotic stability of the error dynamics can still be ensured when $d=0$.

Now, when $d \neq 0$ and $\gamma \neq 0$, then, only UUB stability of the error dynamics can be ensured.
From (\ref{eq:bell}), it can be seen that,
\begin{equation}
\dot{V}=\gamma V-z^TQ_1z-u^TRu+\alpha^2d^Td
\label{v1}
\end{equation}
In order for $\dot{V}$ to be negative, following inequality should hold (using $z^TQ_1z=e^TQe$):
\begin{equation}
\|e\| > \sqrt{\frac{\gamma V^*-\lambda_{min}(R)\|u\|^2+\alpha^2\|d\|^2}{\lambda_{min}(Q)}}
\end{equation}
Provided, $V^*$, $\|u\|$ and $\|d\|$ are finite. 
\end{proof}
As a consequence of Theorem 4 of \cite{modares2015h}, a small value of $\gamma$ and/or positive definite matrix $Q$ (such that its minimum eigenvalue is large) was suggested to ensure asymptotic stability of the error dynamics.

Policy iteration framework is a computation approach to iteratively solve the Bellman equation and improve the control policies. 
It is generally started off with some known initial stabilizing policy $u$ and then following two steps are iteratively repeated till convergence is achieved.

(i) Policy Evaluation: Given initial admissible control and disturbance policies, this step entails solving the Bellman equation as (where $V_i, u_i, d_i$ denote improved value function and policies at $i^{th}$ iteration):
\begin{equation}
\begin{split}
 \nabla{V}_i(F+Gu_i+Kd_i)-\gamma V_i+z^TQ_1z+u_i^TRu_i-\alpha^2d_i^Td_i=0   
\end{split}
\label{iter}
\end{equation}

(ii) Policy Improvement: This step produces improved control and disturbance policies:
\begin{equation}
\begin{split}
u_{i+1}=-\frac{1}{2}R^{-1}G^T\nabla{V}_i ;~~d_{i+1}=\frac{1}{2\alpha^2}K\nabla{V}_i
\end{split}
\label{impro}
\end{equation}

Note that implementation of traditional policy iteration algorithms requires complete knowledge of system dynamics. However, this requirement in policy evaluation step can be obviated via the integral reinforcement learning (IRL) \cite{modares2014optimal}, \cite{vamvoudakis2014online}. In their formulation, requirement of $f(x)$ is precluded, however, $g(x), k(x)$ are still needed to improve the policies. In order to achieve the control completely model-free, Modares and Lewis \cite{modares2015h} presented $H_{\infty}$ tracking control leveraging IRL as well as three NNs - actor, critic and disturbance - to approximate control action, value function and disturbance policy, respectively. 
Motivated by this result, this paper also uses three set of NNs to approximate value function, control and disturbance policies in order to solve $H_{\infty}$-tracking problem with a novel continuous time update law for all three NNs.

\vspace{-.1cm}
\section{Integral Reinforcement Learning \& Value Function Approximation}\label{irl}
\subsection{Derivation of Model-Free HJI equation}\label{model_free_irl}

In order to completely remove the requirement of system dynamics from Policy Evaluation step, the IRL will be utilized in the following way. 
In the subsequent analysis, $u$ refers to the executed control policy and $d$ refers to the disturbance present in the system. 
It is assumed that an initial admissible policy is known.
Improved policies on the other hand are denoted by $u_i,d_i$. 
Then, from (\ref{eq:aug}) the augmented system dynamics can be re-written as,
\begin{equation}
\begin{split}
\dot{z}=F+Gu_i+Kd_i+G(u-u_i)+K(d-d_i)
\end{split}
\label{augaug}
\end{equation}
Taking derivative of $V_i(z)$ along (\ref{augaug}) a revised form of Bellman equation (see (\ref{eq:bell})) is given by,
\begin{equation}
\begin{split}
&V_{zi}(F+Gu_i+Kd_i)+V_{zi}(G(u-u_i))+V_{zi}(K(d-d_i))\\
&-\gamma V_i=-z^TQ_1z-u_i^TRu_i+\alpha^2 d_i^Td_i
\end{split}
\label{infit}
\end{equation}
Multiplying both sides of (\ref{infit}) by $e^{-\gamma t}$,
left hand side (LHS) of (\ref{infit}) can be expressed as,
\begin{equation}
\begin{split}
\frac{d(e^{-\gamma t}V_i(z))}{dt}=e^{-\gamma t}(V_{zi}^T(F+Gu_i+Kd_i)+V_{zi}^T(G(u-u_i))\\
+V_{zi}^T(K(d-d_i))-\gamma V_i)
\end{split}
\label{diffrhs}
\end{equation}
where, $V_{zi}\triangleq \nabla_z{V_i}$. 
Using (\ref{iter}) and (\ref{impro}) in (\ref{diffrhs}), ${d(e^{-\gamma t}V_i)}/{dt}$ can be rewritten as:
\begin{equation}
\begin{split}
\frac{d(e^{-\gamma t}V_i(z))}{dt}&=e^{-\gamma t}(-z^TQ_1z-u_iRu_i+\alpha^2d_i^Td_i\\
&+V_{zi}^TG(u-u_i)+V_{zi}^TK(d-d_i)) \\
&=e^{-\gamma t}(-z^TQ_1z-u_iRu_i+\alpha^2d_i^Td_i\\
&-2u_{i+1}^TR(u-u_i)+2\alpha^2d_{i+1}^T(d-d_i)) 
\end{split}
\label{appr1}
\end{equation}


Integrating the above equation over $[t-T,t]$ on both sides of (\ref{appr1}) and rearranging,
\begin{equation}
\begin{split}
&V_i(t)-e^{\gamma T}V_i(t-T)+\int_{t-T}^{t}e^{-\gamma(\tau-t)}(z^TQ_1z+u_iRu_i\\
&-\alpha^2d_i^Td_i+2u_{i+1}^TR(u-u_i)-2\alpha^2d_{i+1}^T(d-d_i))d\tau=0
\end{split}
\label{model_free_hji}
\end{equation}
where, $V_i(t)\triangleq V_i(z(t))$ and $V_i(t-T)\triangleq V_i(z(t-T))$. Note that (\ref{model_free_hji}) resembles (\ref{infit}) in the limiting sense when $T \to 0$ 
To maintain this equivalence, the reinforcement interval $T$ should be selected as small as possible. However, it is important to note that compared to (\ref{iter}) or (\ref{infit}), equation (\ref{model_free_hji}) does not include system dynamics. 
Eq. (\ref{model_free_hji}) can be rewritten as,
\begin{equation}
\begin{split}
&V_i(t)-e^{\gamma T}V_i(t-T)+I_1+\int_{t-T}^{t}e^{-\gamma(\tau-t)}(2u_{i+1}^TR(u-u_i)\\
&-2\alpha^2d_{i+1}^T(d-d_i))d\tau=0
\end{split}
\label{valint}
\end{equation}
where, 
\begin{equation}
\begin{split}
I_1=\int_{t-T}^{t}e^{-\gamma(\tau-t)}(z^TQ_1z+u_iRu_i-\alpha^2d_i^Td_i)
\end{split}
\label{rein_int}
\end{equation}
\subsection{Approximation of Value function, Control policy and Disturbance policy}\label{val}
Similar to \cite{zhu2016using} and \cite{modares2015h}, value function and improved policies are represented by,
\begin{equation}
\begin{split}
V_i=W_c^T\sigma_c+\varepsilon_c;~u_{i+1}&=W_a^T\sigma_a+\varepsilon_a;~
d_{i+1}=W_d^T\sigma_d+\varepsilon_d
\end{split}
\label{apprx1}
\end{equation}
where, $W_c \in \mathbb{R}^{a_1}$, $\sigma_c \in \mathbb{R}^{a_1}$, $W_a \in \mathbb{R}^{a_2\times m}$, $\sigma_a \in \mathbb{R}^{a_2}$, $W_d \in \mathbb{R}^{a_3\times l}$ and $\sigma_d \in \mathbb{R}^{a_3}$.
Using (\ref{apprx1}) in (\ref{valint}), the HJI error becomes,
\begin{equation}
\begin{split}
\varepsilon_{HJI}=W_c^T[\sigma_c(t)-e^{\gamma T}\sigma_c(t-T)]+I_1\\
+\int_{t-T}^te^{-\gamma(\tau-t)}(2\sigma_a^TW_aR(u-u_i)-2\alpha^2\sigma_d^TW_d(d-d_i))
\end{split}
\end{equation}
Now, since ideal weights are not known, their estimates will be utilized instead,
\begin{equation}
\hat{V}_i=\hat{W}_c^T\sigma_c;~\hat{u}_{i+1}=\hat{W}_a^T\sigma_a;~
\hat{d}_{i+1}=\hat{W}_d^T\sigma_d
\end{equation}
Then, the HJI error in terms of estimated weights can be written as,
\begin{equation}
\begin{split}
\hat{e}_1(t)&=\hat{W}_c^T[\sigma_c(t)-e^{\gamma T}\sigma_c(t-T)]+I_1\\
&+\text{v}(\hat{W}_a)^T\int_{t-T}^t2e^{-\gamma(\tau-t)}(R(u-u_i)\otimes \sigma_a)\\
&-\text{v}(\hat{W}_d)^T\int_{t-T}^t2e^{-\gamma(\tau-t)}(\alpha^2(d-d_i)\otimes\sigma_d)
\end{split}
\label{intgeneral}
\end{equation}
Eq. (\ref{intgeneral}) can be written in compact form as,
\begin{equation}
\begin{split}
\hat{e}_1(t)=\hat{W}^T\rho_1+I_1(t)
\end{split}
\end{equation}
where, $I_1$ is the reinforcement integral given in (\ref{rein_int}). And, 
\begin{equation}
\footnotesize
\begin{split}
\hat{W}=\begin{pmatrix}
\hat{W}_c \\
\text{v}(\hat{W}_a)\\
\text{v}(\hat{W}_d)
\end{pmatrix},\rho_1=\begin{pmatrix}
\Delta\sigma_c \\
\int_{t-T}^t2e^{-\gamma(\tau-t)}(R(u-u_i)\otimes \sigma_a(t))\\
-\int_{t-T}^t2e^{-\gamma(\tau-t)}(\alpha^2(d-d_i)\otimes\sigma_d(t))
\end{pmatrix}
\end{split}
\label{eq:wrho}
\end{equation}
where, $\text{v}(.)$ represents vectorization of matrix and $\otimes$ denotes the kronecker product.
Here, $\hat{W} \in \mathbb{R}^{q}$ is the composite NN weight vector, and $\rho_1 \in \mathbb{R}^q$ is the composite regressor vector, where $q=a_1+ma_2+la_3$, in which $m$ is the dimension of the control vector and $l$ is the dimension of the disturbance vector, and $a_1,a_2,a_3$ are number of neurons in the hidden layer or the size of regressor vector for critic, actor and disturbance, respectively. And, $\Delta\sigma_c=[\sigma_c(t)-e^{\gamma T}\sigma_c(t-T)]$.
In subsequent discussion, $\hat{e}_1(t),~\hat{e}_1(t_j)$ are denoted by $\hat{e}_1$ and $\hat{e}_{1j}$, respectively and $\rho_1 \triangleq \rho_1(t)$, $\rho_{1j} \triangleq \rho_1(t_j)$.
Similarly, $I_1 \triangleq I_1(t)$ and $I_{1j} \triangleq I_{1}(t_j)$.

\subsection{Exisiting update laws in literature for Off-policy IRL}\label{sub:existing}
In \cite{modares2015h}, a recursive least square-based update law was proposed to minimize the HJI approximation error to solve $H_{\infty}$-tracking problem.
Their update law was given by,
\begin{equation}
\hat{W}=(\aleph\aleph^T)^{-1}\aleph Y
\label{eq:modares_upd_law}
\end{equation}
where, $\aleph=[\rho_1(t_1),\rho_1(t_2),\rho_1(t_3),...,\rho_1(t_N)]$ and $Y=[-I_1(t_1)\\,-I_1(t_2),-I_1(t_3),...,-I_1(t_N)]^T$.
This yields $V_i$, $u_{i+1}$ and $d_{i+1}$. 
However, this descrete-time update law requires that $N$ samples be collected, before (\ref{eq:modares_upd_law}) can be implemented.
This procedure make it less sensitive to real time variation in plant parameters. 

A continuous-time update law is more sensitive to real time parametric variations and hence a continuous-time update law was presented in \cite{zhu2016using} utilizing constant learning gradient descent and experience replay (ER) technique to train the actor and critic NN in Off-policy IRL to solve optimal tracking problem. 
A memory stack ($\{\rho_1(t_j),I_1(t_j)\}_{j=1}^N$) containing past observations was constructed, which was then utilized for incorporating the ER terms in the parameter update law in order to improve the performance of gradient descent-based algorithm by iteratively using past observations.
However, this control formulation did not incorporate any disturbance rejection. 
Their update law was given as,
\vspace{-.2cm}
\begin{equation}
\begin{split}
\dot{\hat{W}}=-\frac{\eta}{N+1}\Big(\frac{\rho_2}{m_s^2}\hat{e}_2+\sum_{j=1}^N\Big(\frac{\rho_{2j}}{m_{sj}^2}\hat{e}_{2j}\Big)\Big)
\end{split}
\label{eq:zhu}
\end{equation}
where, $\rho_2$ in (\ref{eq:zhu}) is made up of first two components of $\rho_1$ in (\ref{eq:wrho}), that is $\rho_2$ does not contain the last component of $\rho_1$ that corresponds to disturbance term ($d$).

Recently a continuous time update law was presented in \cite{zhang2017finite} for $H_{\infty}$-tracking control incorporating terminal constraints, in which the update law relied only on constant learning-based gradient descent apart from a term dedicated for incorporating terminal constraints.

\begin{figure}
    \centering
    \includegraphics[width=.52\textwidth,height=9.5cm,keepaspectratio,trim={.9cm 0.0cm .09cm .08cm},clip]{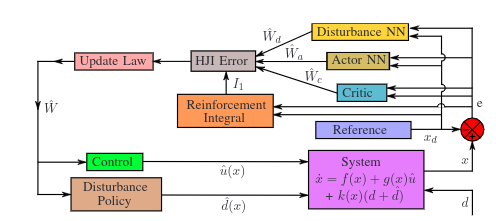}
    \caption{Block diagram of the control system}
    \label{fig:amd1}
\end{figure}

\vspace{-.2cm}
\section{Variable gain gradient descent and experience replay technique-based parameter update law}\label{tuning}
\subsection{Novel update law}
\vspace{-.2cm}
All the update laws mentioned in Section \ref{sub:existing} either utilize recursive least square (RLS) method \cite{modares2015h} or gradient descent with constant learning rate \cite{zhang2017finite}.
While the RLS-based update laws are usually found to be less sensitive to real-time parameter variations in plant dynamics \cite{zhu2016using}, constant learning rate-based update law cannot scale the learning rate based on the instantaneous value of the HJI error \cite{mishra2019variable}. 
Also, experience replay technique and inclusion of robust term in update laws were found to beneficial \cite{zhu2016using,modares2013adaptive}, \cite{liu2015reinforcement}. 
Considering these, a novel continuous time update law is presented in this paper to tune the critic, actor and disturbance NN weights online in order solve $H_{\infty}$-tracking problem.
The novel update law utilizing variable gain gradient descent and ER technique is now  presented as,
\begin{equation}
\begin{split}
\dot{\hat{W}}=-\frac{\eta}{N+1}\Big(\frac{\rho_1}{m_s^2}|\hat{e}_1|^{k_1}\hat{e}_1+\sum_{j=1}^N\Big(\frac{\rho_{1j}}{m_{sj}^2}|\hat{e}_{1j}|^{k_1}\hat{e}_{1j}\Big)\\
-K_1|\hat{e}_1|^{k_1}\frac{\rho_1^T}{m_s}\hat{W}+|\hat{e}_1|^{k_1}K_2\hat{W}-K_1\sum_{j=1}^N\frac{|\hat{e}_{1j}|^{k_1}\rho^T_{1j}}{m_{sj}}\hat{W}\Big)
\end{split}
\label{firstupdate}
\end{equation}
where, $m_s=\sqrt{1+\rho^T_1\rho_1}$ and $m_{sj}=\sqrt{1+\rho^T_{1j}\rho_{1j}}$ and $K_1 \in \mathbb{R}^{q}$ (where $q=a_1+ma_2+la_3$ and $m$ is the dimension of the control vector and $l$ is the dimension of the disturbance vector, and  $a_1,a_2,a_3$ are as defined after (\ref{appr1})) , $K_2 \in \mathbb{R}^{q \times q}$. 

It can be seen that the update law (\ref{firstupdate}) utilizes variable learning rate (via the term $|\hat{e}_1|^{k_1}$) that is a function of instantaneous HJI error.
This has the advantage of scaling the learning rate and reducing the size of the residual set for error in NN weights as will become clear in the stability proof of Theorem \ref{th1}.
Additionally the second and fifth term under summation correspond to the ER terms, these terms can use past observations much more effectively.
The memory stack in ER can be updated with recent data as and when they arrive. 
This leads to an efficient learning from past data. 
Finally, inclusion of robust terms in the update law provides robustness against variations in approximation errors and also reduce the size of the residual set for error in NN weights.

\par
Now, assuming that $u_{i}$ and $d_i$ are in sufficiently small neighborhood of   
the optimal policies, the HJI error can be modified in the same way as mentioned in \cite{zhu2016using} (refer to Section 4.2 in \cite{zhu2016using}). 
It is assumed that there exist NNs that can approximate the optimal value, action and disturbance policies as: 
\begin{equation}
\begin{split}
V^*=W_c^T\sigma_c+\varepsilon_c;~u^*&=W_a^T\sigma_a+\varepsilon_a;~
d^*=W_d^T\sigma_d+\varepsilon_d
\end{split}
\label{appro}
\end{equation}
Under these assumptions, using (\ref{appro}) in (\ref{model_free_hji}), the HJI approximation error is obtained as,
\begin{equation}
\begin{split}
 \varepsilon_{HJI}&=W_c^T[\sigma_c(t)-e^{\gamma T}\sigma(t-T)]+\int_{t-T}^te^{-\gamma(\tau-t)}(z^TQ_1z\\
&+2\sigma_a^TW_aRu-\sigma_a^TW_aRW_a^T\sigma_a+\alpha^2\sigma_d^TW_dRW_d^T\sigma_d\\
&-2\alpha^2\sigma_d^TW_dd)
\end{split}
\label{ehji}
\end{equation}
In terms of approximation errors the HJI error $\varepsilon_{HJI}$ can equivalently be given as,
\begin{equation}
\begin{split}
\varepsilon_{HJI}&=\varepsilon_c(t)-e^{\gamma T}\varepsilon_c({t-T})-\int_{t-T}^t e^{-\gamma(\tau-t)}\Big[2\varepsilon_a^TRu\\
&-2\varepsilon_a^TR\sigma_a-\varepsilon_a^TR\varepsilon_a+\alpha^2\varepsilon_d^TR\varepsilon_d-2\alpha^2\varepsilon_d^Td\Big]
\end{split}
\end{equation}
\begin{assumption}\label{ud}
\textnormal{
It is assumed that the control policy $u$ is admissible policy for the augmented system. 
This makes the augmented system remain in the compact set $\Omega \subset \mathbb{R}^{2n}$.
Such an admissible policy is chosen for online training of critic, actor and disturbance NNs.
}
\end{assumption}

\begin{assumption}\label{norms}
\textnormal{There exist bounds such that, $\|W\| \leq W_M$, $\|\bar{\rho}\| \leq \bar{\rho}_M$, $|m_s| \leq m_{sM}$, $\|W_c\| \leq W_{cm},~\|W_a\| \leq W_{am},~\|W_d\| \leq W_{dm},~\|\sigma_c\| \leq b_c,~\|\sigma_a\| \leq b_a,~\|\sigma_d\| \leq b_d,~\|\varepsilon_c\| \leq b_{\varepsilon c},~\|\varepsilon_a\| \leq b_{\varepsilon a},~\|\varepsilon_d\| \leq b_{\varepsilon d}$. 
This is in line with Assumption 2 of \cite{zhu2016using}}
\end{assumption}

Now, since, ideal NN weights are not known, their estimates will be utilized to express optimal value function and optimal policies.
\begin{equation}
\begin{split}
\hat{V}^*=\hat{W}_c\sigma_c,~\hat{u}^*=\hat{W}_a\sigma_a,~\hat{d}^*=\hat{W}_d\sigma_d 
\end{split}
\end{equation}
Based on these estimated weights, following (Eq. (\ref{ehji})) the approximate HJI error can be re-stated as:
\begin{equation}
\begin{split}
\hat{e}&=\hat{W}_c^T[\sigma_c(t)-e^{\gamma T}\sigma(t-T)]+\int_{t-T}^te^{-\gamma(\tau-t)}(z^TQ_1z\\
&+2\sigma_a^T\hat{W}_aRu-\sigma_a^T\hat{W}_aR\hat{W}_a^T\sigma_a+\alpha^2\sigma_d^T\hat{W}_dR\hat{W}_d^T\sigma_d\\
&-2\alpha^2\sigma_d^T\hat{W}_dd)
\end{split}
\end{equation}
The approximate HJI error, thus, can be expressed in a compact form as:
\begin{equation}
\begin{split}
\hat{e}(t)&=\hat{W}^T\rho+\text{v}(\hat{W}_a)^T\mathcal{A}_2\text{v}(\hat{W}_a)-\text{v}(\hat{W}_d)^T\mathcal{B}_2\text{v}(\hat{W}_d)+I_2
\end{split}
\label{e}
\end{equation}
where, 
\begin{equation}
\begin{split}
\Delta\sigma_c&\triangleq \sigma_c(t)-e^{\gamma T}\sigma(t-T) \\ \mathcal{A}_1&\triangleq\int_{t-T}^te^{-\gamma(\tau-t)}(Ru\otimes\sigma_a)d\tau \\
\mathcal{A}_2&\triangleq\int_{t-T}^te^{-\gamma(\tau-t)}(R\otimes\sigma_a\sigma_a^T)d\tau\\
\mathcal{B}_1&\triangleq\int_{t-T}^te^{-\gamma(\tau-t)}(\alpha^2d\otimes\sigma_d)d\tau \\
\mathcal{B}_2&\triangleq\int_{t-T}^te^{-\gamma(\tau-t)}(\alpha^2\otimes\sigma_d\sigma_d^T)d\tau\\
I_2&\triangleq\int_{t-T}^te^{-\gamma(\tau-t)}(z^TQ_1z)d\tau
\end{split}
\label{a2b2}
\end{equation}

\begin{equation}
\begin{split}
\hat{W}\triangleq\begin{pmatrix}
\hat{W}_c \\
\text{v}(\hat{W}_a) \\ 
\text{v}(\hat{W}_d)
\end{pmatrix},~\rho\triangleq\begin{pmatrix}
\Delta\sigma_c \\
2\mathcal{A}_1-2\mathcal{A}_2\text{v}(\hat{W}_a) \\
-2\mathcal{B}_1+2\mathcal{B}_2\text{v}(\hat{W}_d)
\end{pmatrix}
\end{split}
\label{wrho}
\end{equation}
Now, in order to implement experience replay technique, past data $\left(\{\hat{e}(t_j),m_s(t_j),\rho(t_j),I_2(t_j)\}_{j=1}^N\right)$ is stored in memory stack of size $N$. These data that are stored in memory stack are given as,
\begin{equation}
\begin{split}
\rho(t_j)&\triangleq\begin{pmatrix}
\Delta\sigma_c(t_j) \\
2\mathcal{A}_1(t_j)-2\mathcal{A}_2(t_j)\text{v}(\hat{W}_a) \\
-2\mathcal{B}_1(t_j)+2\mathcal{B}_2(t_j)\text{v}(\hat{W}_d)
\end{pmatrix} \\
m_s(t_j)&\triangleq\sqrt{1+\rho^T(t_j)\rho(t_j)} \\
\hat{e}(t_j)&\triangleq\hat{W}^T\rho(t_j)+\text{v}(\hat{W}_a)^T\mathcal{A}_2(t_j)\text{v}(\hat{W}_a)\\
&-\text{v}(\hat{W}_d)^T\mathcal{B}_2(t_j)\text{v}(\hat{W}_d)+I_2(t_j)
\end{split}
\end{equation}
where, $m_s=\sqrt{1+\rho^T\rho}$. Also in the subsequent analysis, $\rho_j \triangleq \rho(t_j),  m_{sj} \triangleq m_s(t_j),  \hat{e}_j\triangleq \hat{e}(t_j)$ for $j=1,2,3...N$.
\par
Similar to (\ref{firstupdate}), continuous-time update law for this case can be written as, 
\begin{equation}
\begin{split}
\dot{\hat{W}}=-\frac{\eta}{N+1}\Big(\frac{\rho}{m_s^2}|\hat{e}|^{k_1}\hat{e}+\sum_{j=1}^N\Big(\frac{\rho_j}{m_{sj}^2}|\hat{e}_j|^{k_1}\hat{e}_j\Big)\\
-K_1|\hat{e}|^{k_1}\frac{\rho^T}{m_s}\hat{W}+|\hat{e}|^{k_1}K_2\hat{W}-K_1\sum_{j=1}^N\frac{|\hat{e}_j|^{k_1}\rho^T(t_j)}{m_{sj}}\hat{W}\Big)
\end{split}
\label{upd_law}
\end{equation}
It could be observed that certain terms appearing in (\ref{upd_law}) are defined differently from (\ref{firstupdate}).
For instance, ($\hat{e}$ and $\rho$) appearing in (\ref{upd_law}) are given by (\ref{e}) and (\ref{wrho}), respectively. 

Note that the update law presented above (\ref{upd_law}), is different from least square-based update law mentioned in \cite{modares2015h} and continuous time gradient descent-based update laws mentioned in \cite{zhu2016using} and \cite{zhang2017finite}.
The update law presented in this paper consists of five terms. The first term is directly responsible for reducing the HJI error, while the second term is a representation of its past observations over the memory stack.  
Also, unlike the constant learning rate in \cite{zhu2016using} and \cite{zhang2017finite}, in this paper the learning rate in (\ref{upd_law}) is time-varying and considered as a function of the HJI error such that it can accelerate the learning when the HJI error is large and reduce the learning speed when the HJI error becomes small. 
The next three terms are responsible for providing robustness in achieving small residual set.
Moreover, the second and fifth term correspond to the experience replay (ER) of the first and third terms, respectively. Significance of each term in the update law (\ref{upd_law}) in improving the performance of the tracking controller would be evident in the proof of Theorem \ref{th1} and its subsequent discussion.

\subsection{Stability proof of the update law}
\begin{theorem}\label{th1}
Let $\hat{W}$ be the estimated parameters for critic, actor and disturbance. Under the Assumptions \ref{ud}, \ref{norms}, and that the normalized regressor $\bar{\rho}\triangleq\rho/\sqrt{1+\rho^T\rho}$ is persistently excited, the update law mentioned in (\ref{upd_law}) ensures the error in NN weights $\tilde{W}$ to be UUB stable.
\end{theorem}
\begin{proof}
Let the Lyapunov function candidate be: $L=(1/2)\tilde{W}^T\eta^{-1}\Tilde{W}$. In order to prove stability of the update law, the HJI error needs to be expressed as a function of $\tilde{W}$. 
In order to accomplish this, using (\ref{ehji}) and (\ref{e}),
\begin{equation}
\begin{split}
 \hat{e}&=\varepsilon_{HJI}-\tilde{W}^T\rho-\text{v}(\tilde{W}_a)^T\mathcal{A}_2\text{v}(\tilde{W}_a)+\text{v}(\tilde{W}_d)^T\mathcal{B}_2\text{v}(\tilde{W}_d) 
\end{split}
\end{equation}

In subsequent discussion, $\varepsilon$ would be equivalently used in place of $\varepsilon_{HJI}$. Differentiating the Lyapunov function,

\begin{equation}
\begin{split}
\dot{L}&=\tilde{W}^T\eta^{-1}\dot{\tilde{W}}=\frac{g_1}{(N+1)m_s^2}(\tilde{W}^T\rho\varepsilon-\tilde{W}^T\rho\tilde{W}^T\rho\\
&-\tilde{W}^T\rho\text{v}(\tilde{W}_a)^T\mathcal{A}_2\text{v}(\tilde{W}_a)+\tilde{W}^T\rho\text{v}(\tilde{W}_d)^T\mathcal{B}_2\text{v}(\tilde{W}_d))\\
&+\frac{1}{N+1}\Big(\tilde{W}\varepsilon\sum_{j=1}^N\frac{\rho(t_j)g_1(t_j)}{m_s^2(t_j)}-\tilde{W}^T\sum_{j=1}^N\frac{\rho(t_j)\rho(t_j)^Tg_1(t_j)}{m_s^2(t_j)}\tilde{W}\\
&-\tilde{W}^T\sum_{j=1}^N\frac{g_1(t_j)\rho(t_j)^T}{m_s^2(t_j)}\text{v}(\tilde{W}_a)^T\mathcal{A}_2(t_j)\text{v}(\tilde{W}_a)\\
&+\tilde{W}^T\sum_{j=1}^N\frac{\rho(t_j)g_1(t_j)}{m_s^2(t_j)}(\text{v}(\tilde{W}_d)^T\mathcal{B}_2(t_j)\text{v}(\tilde{W}_d))\Big)\\
&-\frac{1}{N+1}g_1\tilde{W}^TK_1^T\frac{\rho^T}{m_s}W+\frac{g_1}{N+1}\tilde{W}^TK_1^T\frac{\rho^T}{m_s}\tilde{W}\\
&+\frac{g_1}{N+1}\tilde{W}^TK_2W-\frac{g_1}{N+1}\tilde{W}^TK_2\tilde{W}\\
&-\tilde{W}^TK_1\sum_{j=1}^N\frac{g_1(t_j)\rho(t_j)^T}{m_s(N+1)}W+\tilde{W}^TK_1\sum_{j=1}^N\frac{g_1(t_j)\rho(t_j)^T}{m_s(N+1)}\tilde{W}
\end{split}
\label{firsteqL}
\end{equation}
where, 
\begin{equation}
\begin{split}
  g_1=|\hat{e}|^{k_1} ,~g_1(t_j)=|\hat{e}_j|^{k_1},~(j=1,2,...,N) 
\end{split}
\label{g1g2}
\end{equation}

Now, in order to find a bound over $\dot{L}$, it is required to find bound over terms containing $\mathcal{A}_2$ and $\mathcal{B}_2$ (ref. to Eq. (\ref{a2b2})). 
Recall from Assumption \ref{norms} that $\|\sigma_a\| \leq b_a$ and $\|\sigma_d\| \leq b_d$.
Utilizing properties of matrices (maximum and minimum eigenvalue) and persistent excitation (PE) condition ($\lambda_1 I \leq \int_{t-T}^t\rho\rho^Td\tau \leq \lambda_2 I$, where $\lambda_1,\lambda_2$ are positive constants) on regressor, the bounds over terms containing $\mathcal{A}_2$ and $\mathcal{B}_2$ can be derived as,
\begin{equation}
\begin{split}
\text{v}(\tilde{W}_a)(R\otimes\sigma_a\sigma_a^T)\text{v}(\tilde{W}_a)\leq q_1\|\text{v}(\tilde{W}_a)\|^2 \leq q_1\|\tilde{W}\|^2\\
\text{v}(\tilde{W}_d)(\alpha^2\otimes\sigma_d\sigma_d^T)\text{v}(\tilde{W}_d)\leq q_2\|\text{v}(\tilde{W}_d)\|^2 \leq q_2\|\tilde{W}\|^2
\end{split}
\label{a2matrix}
\end{equation}
where, $q_1$ and $q_2$ are the maximum eigenvalues of the matrices given by $
(R\otimes\sigma_a\sigma_a^T)$ and $(\alpha^2\otimes\sigma_d\sigma_d^T)$, respectively. Further, the bound over terms, $\text{v}(\tilde{W}_a)\mathcal{A}_2\text{v}(\tilde{W}_a)$ and $\text{v}(\tilde{W}_d)\mathcal{B}_2\text{v}(\tilde{W}_d)$ can be derived from (\ref{a2b2}), (\ref{wrho}) and (\ref{a2matrix}) as,

\begin{equation}
\begin{split}
\text{v}(\tilde{W}_a)&\mathcal{A}_2\text{v}(\tilde{W}_a)\leq q_1\int_{t-T}^te^{-\gamma(\tau-t)}\|\tilde{W}\|^2d\tau \\
&\leq \frac{q_1}{\gamma\beta_1}(e^{\gamma T}-1)\|\tilde{W}^T\bar{\rho}\|^2=\frac{q_1}{\gamma\beta_1}(e^{\gamma T}-1)\tilde{W}^T\bar{\rho}\bar{\rho}^T\tilde{W}
\end{split}
\label{va}
\end{equation}
\begin{equation}
\begin{split}
\text{v}(\tilde{W}_d)&\mathcal{B}_2\text{v}(\tilde{W}_d)\leq q_2 \int_{t-T}^te^{-\gamma(\tau-t)}\|\tilde{W}\|^2d\tau \\
&\leq \frac{q_2}{\gamma\beta_1}(e^{\gamma T}-1)\|\tilde{W}^T\bar{\rho}\bar\|^2= \frac{q_2}{\gamma\beta_1}(e^{\gamma T}-1)\tilde{W}^T\bar{\rho}\bar{\rho}^T\tilde{W}
\end{split}
\label{vd}
\end{equation}
where, $\beta_1=\|\rho\|^2$. 
Using the same PE condition on $\mathcal{A}_2,\mathcal{B}_2$ in $\rho$, from (\ref{wrho}) there exists a constant $L_1$ such that,
\begin{equation}
\begin{split}
\Big|\tilde{W}^T\frac{\rho}{m_s^2}\Big| \leq L_1\|\tilde{W}\|
\end{split}
\label{w}
\end{equation}
Combining (\ref{va}), (\ref{vd}) and (\ref{w}),
\begin{equation}
\begin{split}
\Big|\tilde{W}^T\frac{\rho}{m_s^2}\text{v}(\tilde{W}_a)\mathcal{A}_2\text{v}(\tilde{W}_a)\Big| & \leq L_1\frac{q_1}{\gamma\beta_1}(e^{\gamma T}-1)\|\tilde{W}\|\tilde{W}^T\bar{\rho}\bar{\rho}^T\tilde{W} \\
\Big|\tilde{W}^T\frac{\rho}{m_s^2}\text{v}(\tilde{W}_d)\mathcal{B}_2\text{v}(\tilde{W}_d)\Big| & \leq L_1\frac{q_2}{\gamma\beta_1}(e^{\gamma T}-1)\|\tilde{W}\|\tilde{W}^T\bar{\rho}\bar{\rho}^T\tilde{W}
\end{split}
\label{curL}
\end{equation}
where, $\bar{\rho}=\rho/m_s$ and
the reinforcement interval $T$ can be selected such that,
\begin{equation}
\begin{split}
L_1\frac{q_1}{\gamma\beta_1}(e^{\gamma T}-1)\|\tilde{W}\| \leq \varepsilon_{Ta} \\
L_1\frac{q_2}{\gamma\beta_2}(e^{\gamma T}-1)\|\tilde{W}\|\leq \varepsilon_{Td}
\end{split}
\label{eta}
\end{equation}
where, $\varepsilon_{Ta}$ and $\varepsilon_{Td}$ are two small positive scalar constants. 
Therefore, using (\ref{eta}) in (\ref{curL}),
\begin{equation}
\begin{split}
\Big|\tilde{W}^T\frac{\rho}{m_s^2}\text{v}(\tilde{W}_a)\mathcal{A}_2\text{v}(\tilde{W}_a)\Big|&\leq \varepsilon_{Ta}\tilde{W}^T\bar{\rho}\bar{\rho}^T\tilde{W}\\
\Big|\tilde{W}^T\frac{\rho}{m_s^2}\text{v}(\tilde{W}_d)\mathcal{B}_2\text{v}(\tilde{W}_d)\Big|&\leq \varepsilon_{Td}\tilde{W}^T\bar{\rho}\bar{\rho}^T\tilde{W}\\
\end{split}
\label{2term}
\end{equation}
Similarly, their ER versions can be represented as,
\begin{equation}
\small
\begin{split}
\Big|\tilde{W}^T\sum_{j=1}^N\frac{g_{1j}\rho(t_j)}{m_s^2(t_j)}\text{v}(\tilde{W}_a)\mathcal{A}_2(t_j)\text{v}(\tilde{W}_a)\Big| \leq \varepsilon_{Tsa}\tilde{W}^T\sum_{j=1}^N\bar{\rho}_j\bar{\rho}_j^T\tilde{W}\\
\Big|\tilde{W}^T\sum_{j=1}^N\frac{g_{1j}\rho(t_j)}{m_s^2(t_j)}\text{v}(\tilde{W}_a)\mathcal{B}_2(t_j)\text{v}(\tilde{W}_a)\Big| \leq \varepsilon_{Tsd}\tilde{W}^T\sum_{j=1}^N\bar{\rho}_j\bar{\rho}_j^T\tilde{W}
\end{split}
\label{erab}
\end{equation}
Now, using (\ref{2term}) and (\ref{erab}), Eq. (\ref{firsteqL}) can be rewritten as:
\begin{equation}
\small
\begin{split}
&\dot{L} \leq \tilde{W}^T\varepsilon\Big(\frac{\rho g_1}{m_s^2(N+1)}+\sum_{j=1}^N\frac{\rho(t_j)g_1(t_j)}{m_s^2(t_j)(N+1)}\Big)\\
&-\tilde{W}^T\Big(\frac{g_1\bar{\rho}\bar{\rho}^T}{(N+1)}+\sum_{j=1}^N\frac{g_{1j}\bar{\rho}_{j}\bar{\rho}_{j}^T}{(N+1)}\Big)\tilde{W}\\
&+\tilde{W}^T\Big(\frac{1}{N+1}\varepsilon_{Ta}\bar{\rho}\bar{\rho}^T+\frac{1}{N+1}\varepsilon_{Tsa}\sum_{j=1}^N\bar{\rho}_j\bar{\rho}_j^T\Big)\tilde{W}\\
&+\tilde{W}^T\Big(\frac{1}{N+1}\varepsilon_{Td}\bar{\rho}\bar{\rho}^T+\frac{1}{N+1}\varepsilon_{Tsd}\sum_{j=1}^N\bar{\rho}_j\bar{\rho}_j^T\Big)\tilde{W}\\
&-\frac{1}{N+1}g_1\tilde{W}^TK_1^T\frac{\rho^T}{m_s}W+\frac{g_1}{N+1}\tilde{W}^TK_1^T\frac{\rho^T}{m_s}\tilde{W}\\
&+\frac{g_1}{N+1}\tilde{W}^TK_2W-\frac{g_1}{N+1}\tilde{W}^TK_2\tilde{W}\\
&-\tilde{W}^TK_1\sum_{j=1}^N\frac{g_1(t_j)\rho(t_j)^T}{m_s(N+1)}W+\tilde{W}^TK_1\sum_{j=1}^N\frac{g_1(t_j)\rho(t_j)^T}{m_s(N+1)}\tilde{W}
\end{split}
\label{final_inequ}
\end{equation}

After further simplification, (\ref{final_inequ})  can be rendered into the following inequality:
\begin{equation}
\begin{split}
\dot{L}&\leq \frac{-\mathcal{P}^TM\mathcal{P}+\mathcal{P}\mathcal{N}-\mathcal{P}^TM_{\varepsilon_{Ta}}\mathcal{P}+\mathcal{P}^TM_{\varepsilon_{Td}}\mathcal{P} }{N+1}
\end{split}
\label{ldot}
\end{equation}
where,
\begin{equation}
\begin{split}
\mathcal{P}&\triangleq\begin{pmatrix}
\tilde{W}^T\bar{\rho},~
\tilde{W},~\tilde{W}^T\sum_{j=1}^N\bar{\rho}(t_j)
\end{pmatrix}^T\\
M&\triangleq\begin{pmatrix}
g_1 & -\frac{g_1K_1^T}{2} & \sum{g_{1j}} \\
-\frac{g_1K_1}{2} & g_1K_2 & -K_1\frac{\sum{g_{1j}}}{2} \\
-\sum{g_{1j}} & -K_1^T\frac{\sum{g_{1j}}}{2} & \sum{g_{1j}}
\end{pmatrix}\\
\mathcal{N}&\triangleq\begin{pmatrix}
g_1\varepsilon-g_1K_1^TW \\
g_1K_2W\\
\varepsilon\sum_{j=1}^Ng_{1j}-K_1\sum_{j=1}^Ng_{1j}W
\end{pmatrix}
\end{split}
\label{negdef2}
\end{equation}
\begin{equation}
\small
\begin{split}
M_{\varepsilon_{Ta}}\triangleq\begin{pmatrix}
\varepsilon_{Ta} & G_1^T  &  -c_1 \\
-G_1              &  0_{q\times q}     &   L_1 \\
c_1             &  -L_1^T  &  \varepsilon_{Tsa}
\end{pmatrix};M_{\varepsilon_{Td}}\triangleq\begin{pmatrix}
\varepsilon_{Td} & -G_2^T  &  c_2 \\
G_2              &  0_{q\times q}     &   -L_2 \\
-c_2             &  L_2^T  &  \varepsilon_{Tsd}
\end{pmatrix}
\end{split}
\label{metad}
\end{equation}
where, $K_1 \in \mathbb{R}^q$ and $K_2\in\mathbb{R}^{q\times q}$ and $q$ being the dimension of the composite regressor vector $\rho$ (see (\ref{wrho})). Also, $G_1,G_2 \in \mathbb{R}^q$, $c_1,c_2 \in \mathbb{R}$, $L_1,L_2 \in \mathbb{R}^q$ are constants used in (\ref{metad}). 
\begin{prop}\label{pro}
Let $x \in \mathbb{R}^n$ and $M \in \mathbb{R}^{n\times n}$ be any square matrix, then, $\lambda_{min}(\frac{M+M^T}{2})\|x\|^2 \leq x^TMx \leq \lambda_{max}(\frac{M+M^T}{2})\|x\|^2$.
Where, $\lambda_{min}(.)$ and $\lambda_{max}(.)$ denote the minimum and maximum eigen values of corresponding matrices, respectively.
\end{prop}
\begin{proof}
The proof of this proposition is provided in Lemma \ref{mm} in Section \ref{sec14}.
\end{proof}

Using Proposition \ref{pro}, (\ref{ldot}) can be simplified into:
\begin{equation}
\begin{split}
\dot{L} & \leq (-\lambda_{min}(M^{'})\|\mathcal{P}\|^2+b_N\|\mathcal{P}\|-\lambda_{min}(M^{'}_{\varepsilon_{Ta}})\|\mathcal{P}\|^2\\
&+\lambda_{max}(M^{'}_{\varepsilon_{Td}})\|\mathcal{P}\|^2)/(N+1)
\end{split}
\label{ldm}
\end{equation}
where, $b_N$ is the maximum value of norm of $\mathcal{N}$, i.e, $\|\mathcal{N}\| \leq b_N$ and is expressed using (\ref{negdef2}) and Assumptions \ref{norms} as,
\begin{equation}
\begin{split}
b_N&=\Big((g_1\varepsilon-g_1\|K_1\|\|W\|)^2+(g_1\|K_2\|\|W\|)^2+...\\
&(\varepsilon\sum_{j=1}^Ng_{1j}-K_1\sum_{j=1}g_{1j}\|W\|))^2\Big)^{\frac{1}{2}}
\end{split}
\label{BN}
\end{equation}
Note that in (\ref{ldm}), following substitutions were made,
\begin{equation}
\begin{split}
M^{'}_{\varepsilon_{Ta}}& \triangleq\frac{M_{\varepsilon_{Ta}}+M^T_{\varepsilon_{Ta}}}{2},~M^{'}_{\varepsilon_{Td}} \triangleq \frac{M_{\varepsilon_{Td}}+M^T_{\varepsilon_{Td}}}{2}\\
M^{'}& \triangleq \frac{M+M^T}{2}
\end{split}
\end{equation}
From (\ref{ldm}), in order to ensure negative definiteness of $\dot{L}$, the following inequality should hold:
\begin{equation}
\begin{split}
&\|\mathcal{P}\| > \frac{b_N}{\lambda_{min}(M^{'})+\lambda_{min}(M^{'}_{\varepsilon_{Ta}})-\lambda_{max}(M^{'}_{\varepsilon_{Td}})}
\\ \implies&\dot{L}<0
\end{split}
\label{negdef}
\end{equation}
From the definition of $\mathcal{P}$ as mentioned in (\ref{negdef2}), 
\begin{equation}
\begin{split}
\|\mathcal{P}\| \leq \|\tilde{W}\|\sqrt{1+\|\bar{\rho}_M\|^2+\Big\|\sum_{j=1}^N\bar{\rho}_M(t_j)\Big\|^2}
\end{split}
\label{ymax}
\end{equation}
The right hand side of (\ref{ymax}) will be represented as,
\begin{equation}
\mathcal{S}\triangleq\sqrt{1+\|\bar{\rho}_M\|^2+\Big\|\sum_{j=1}^N\bar{\rho}_M(t_j)\Big\|^2}
\label{S}
\end{equation}
From (\ref{negdef}), (\ref{ymax}) and (\ref{S}) the UUB set for error in NN weights is obtained as,
\begin{equation}
\begin{split}
\|\tilde{W}\| > \frac{b_N}{\mathcal{S}(\lambda_{min}(M^{'})+\lambda_{min}(M^{'}_{\varepsilon_{Ta}})-\lambda_{max}(M^{'}_{\varepsilon_{Td}}))}
\end{split}
\label{w_lim}
\end{equation}
Thus, from (\ref{negdef}), (\ref{ymax}) and (\ref{w_lim}), under the NN parameter update law (\ref{upd_law}), the error in NN weights are guaranteed to decrease outside the residual ball given as,
\begin{equation}
\footnotesize
\begin{split}
\Omega_{\tilde{W}}=\Bigg\{\tilde{W}:\|\tilde{W}\| \leq \frac{b_N}{\mathcal{S}(\lambda_{min}(M^{'})+\lambda_{min}(M^{'}_{\varepsilon_{Ta}})-\lambda_{max}(M^{'}_{\varepsilon_{Td}}))}\Bigg\}
\end{split}
\label{ball}
\end{equation}
This concludes the stability proof of the continuous-time update mechanism. 
\end{proof}

\subsection{Discussion on the presented update law}
Note that the update law presented in (\ref{upd_law}) is different from the gradient descent-based update laws of \cite{zhu2016using} and \cite{zhang2017finite} and least square-based one presented in \cite{modares2015h} in several ways. First of all, being a continuous-time update law based on gradient descent, it is more sensitive to variations in plant dynamics than least square-based update mechanism in \cite{modares2015h}. 
Secondly, unlike \cite{zhu2016using}, it utilizes $H_{\infty}$ framework for disturbance rejection as well. 
While \cite{zhang2017finite} utilized $H_{\infty}$ framework for their tracking controller, their gradient descent had only constant learning rate and lacked ER and robust terms to further shrink the size of the residual set.
The prime novelties of the update law (\ref{upd_law}) are the use of variable gain gradient descent and incorporation of robust terms i.e., the last three terms in (\ref{upd_law}). 
These help in improving the performance of the final learnt control policies to track a given reference trajectory.

From Theorem \ref{th1} it is evident that $\|\tilde{W}\|$ decreases in the stable region, i.e., where $\dot{L}$ is negative definite. 
This results in estimated NN weights, i.e., $\hat{W}$ getting closer to ideal NN weights $W$, which in turn implies that the HJI error (\ref{g1g2}) is decreasing in the stable region.
Now, note that the numerator in the right hand side (RHS) of (\ref{ball})) i.e $b_N$ is a function of $g_1=|\hat{e}|^{k_1}$ and $g_{1j}=|\hat{e}(t_j)|^{k_1}$ (see (\ref{BN})), which implies that the size of the ball (\ref{ball}) shrinks due to decreasing $g_1$ and $g_{1j}$.

Thus, $b_N$ encapsulates the effect of variable gain gradient descent in Off-Policy parameter update law.
Further, the variable gain in gradient descent, i.e., $|\hat{e}|^{k_1}$ and  $|\hat{e}_j|^{k_1}, j=1,2,...,N$ scale the learning rate based on instantaneous and past values of HJI error, respectively, where the constant $k_1 \geq 0$ governs the amount of scaling in the learning of gradient descent. 
These terms increase the learning rate when the HJI error is large and slow it down as the HJI error becomes smaller in magnitude. 
So, the actual learning rate becomes, $l=\eta|\hat{e}|^{k_1}$. 
Note that if the $|\hat{e}| \leq 1$, then $l \leq \eta$ for all $k_1 \geq 0$. However, if $|\hat{e}| \geq 1$, then $l \geq \eta$ for all $k_1 \geq 0$


Furthermore, the gains $K_1$ and $K_2$ in the robust term of the adaptation law (\ref{upd_law}) can be selected, so as to have a large $\lambda_{min}((M+M^T)/2)$ (ref. to (\ref{negdef2})), which in turn leads to a smaller ball (ref. to (\ref{ball})) and hence a tighter residual set for $\tilde{W}$. With these novel modifications, the variable gain gradient descent-based Off-policy update law presented in this paper yields a much tighter residual set for $\tilde{W}$ and hence improved tracking performance.


\section{Simulation Results}\label{res}

In order to evaluate the performance of update law proposed in this paper, two applications are considered for simulation studies in this section.
\begin{itemize}
    \item Nonlinear system \cite{zhu2016using} in subection \ref{nl} 
    \item Linearized F16 Model \cite{modares2015h} in subsection \ref{lin}
\end{itemize}

\subsection{A Nonlinear System}\label{nl}
Dynamics of a nonlinear system is considered from \cite{zhu2016using} and is described as,
\begin{equation}
\begin{split}
\dot{x}_1&=-\sin{x_1}+x_2 \\
\dot{x}_2&=-x_1^3+u+d \\
y&=x_1
\end{split}
\label{nlsys}
\end{equation}
where, disturbance affecting the system is given as $d=.1e^{-.1t}\sin{.1t}$. The reference system is considered as \cite{zhu2016using},
\begin{equation}
\begin{split}
\dot{x}_{d}=\begin{pmatrix}
0 & 0.3 \\
-0.3 & 0
\end{pmatrix}\begin{pmatrix}
0.1\sin(0.3t) \\
0.1\sin(0.3t)
\end{pmatrix}
\end{split}
\end{equation}
The penalty on states as appearing in (\ref{l2_perf}), i.e., $Q_1$ is chosen to be:
\begin{equation}
Q_1=\begin{pmatrix}
diag(217,0) & 0 \\
0  & 0
\end{pmatrix}
\end{equation}
Here, $x_{d1}$ is the desired trajectory for the output $y=x_1$. The inital state of the system is $(.5;.5)$. The regressor vectors for critic, actor and disturbance NNs are chosen as :
\begin{equation}
\begin{split}
\sigma_c&=\begin{pmatrix}
z_1^2,~z_2^2,~z_3^2,~z_4^2,~z_1z_2,~z_1z_3,~z_1z_4,~z_2z_3,~z_2z_4,~z_3z_4
\end{pmatrix}\\
\sigma_a&=\begin{pmatrix}
z_1,~z_2,~z_3,~z_4
\end{pmatrix}\\
\sigma_d&=\begin{pmatrix}
z_1^2,~z_2^2,~z_1z_3,~z_1z_4,~z_1z_2
\end{pmatrix}
\end{split}
\end{equation}
where, $z=(e^T,x_d^T)^T \in \mathbb{R}^4$ and $z_i$ is individual component of $z$. 
The constant part of the learning rate for both the cases is selected to be, $\eta=2998$, the size of memory stack, i.e., $N$ for experience replay technique is chosen to be $20$. 
The level of attenuation $\alpha$ is choosen to be $0.01$. 
The value of reinforcement interval should be selected as small as possible in order to preserve the relationship between Bellman equation and IRL equation (refer to section \ref{model_free_irl}).
Here, the reinforcement interval $T$ is selected as 0.001s. All the NN weights are initialized to $0$. 
Also, note that in order to yield tighter residual set, $\lambda_{min}(M')$ in the denominator of RHS of  (\ref{ball}) needs to be large, which could be made possible by selecting the gains in robust term $K_1$ and $K_2$ with high norms. However, it should also be noted that since norms of both $K_1$ and $K_2$ appear in numerator of RHS of  (\ref{ball}) too. Hence, gains $K_1$ and $K_2$ cannot be selected with very high norms. For ease in the simulation study, $K_1$ and $K_2$ are both selected as $0_q$ and ($0_{q\times q}$) (for both the cases), where $q$ is the dimension of composite regressor vector $\rho$ (refer to (\ref{wrho})), i.e., $q=19$.
At first, an exploratory control policy is fired into the system, and the system is allowed to explore the state space. The exploratory control signal considered has the form, $n(t)=2e^{-0.009t}\big(\sin(11.9t)^2\cos(19.5t)+\sin(2.2t)^2\cos(5.8t)+\sin(1.2t)^2\cos(9.5t)+\sin(2.4t)^5\big)$ similar to the one mentioned in \cite{vamvoudakis2014online}.
When critic, actor and disturbance NN weights have converged, the exploration is stopped and learnt policies are executed to the system as shown in Fig. \ref{fig:amd1}.

\begin{figure*}
\centering
\subcaptionbox{Critic NN weights\label{fig:critic}}{\includegraphics[width=.32\textwidth,height=9.5cm,keepaspectratio,trim={1.8cm 0.0cm 4cm .08cm},clip]{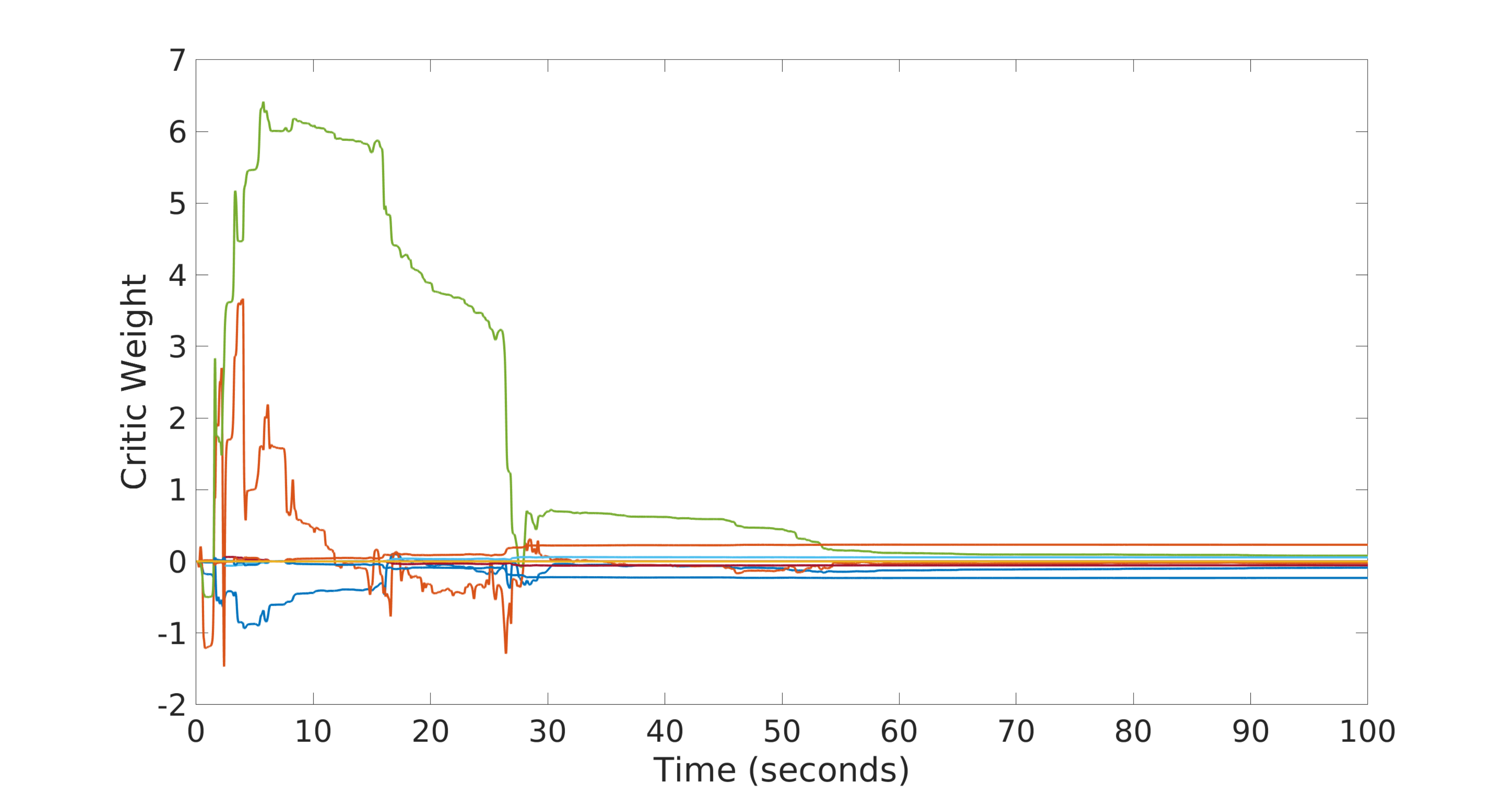}}%
\hspace{0cm} 
\subcaptionbox{Actor NN weights\label{fig:actor}}{\includegraphics[width=.32\textwidth,height=9.5cm,keepaspectratio,trim={1.8cm 0.0cm 2cm .08cm},clip]{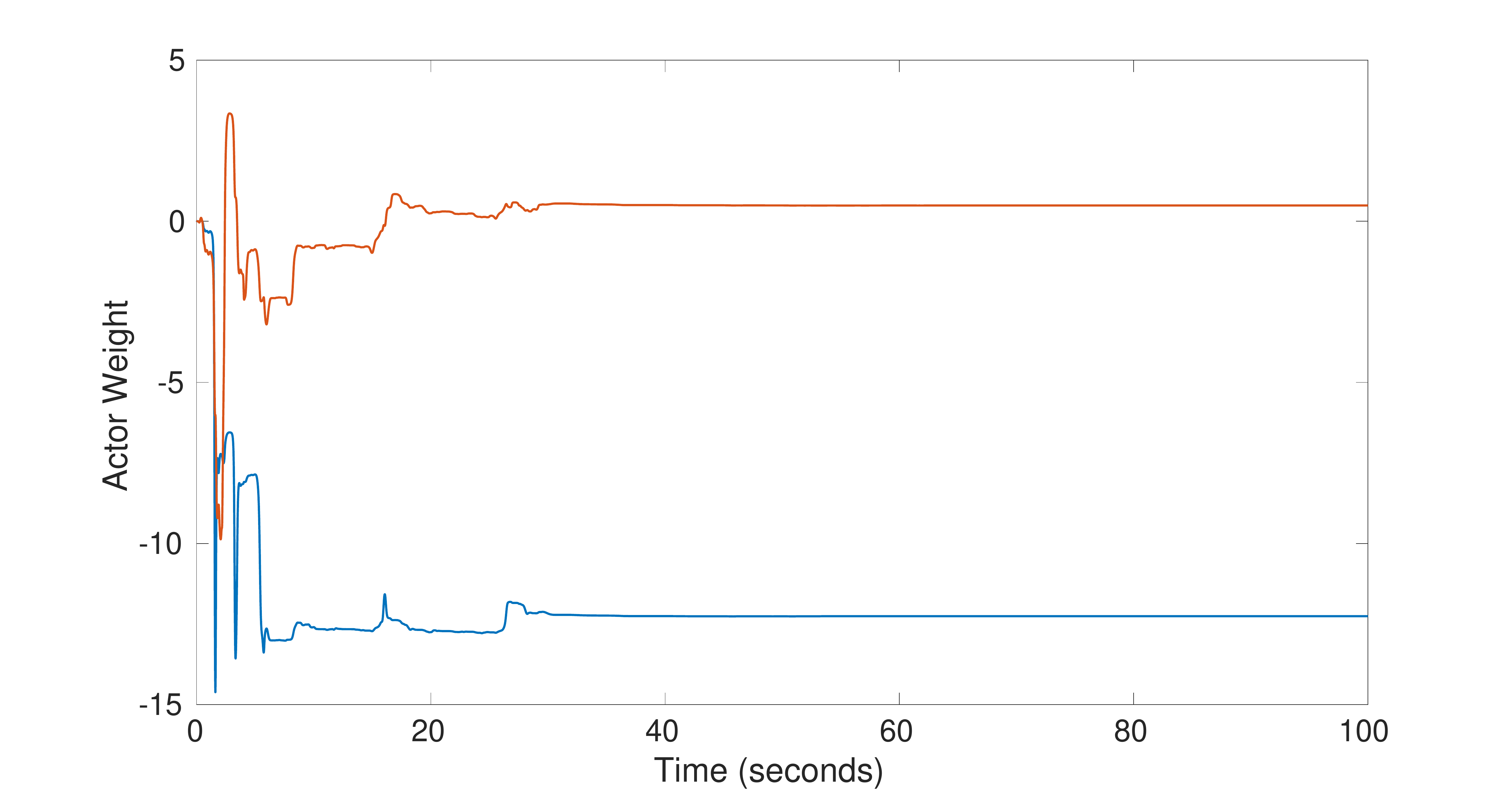}}
\hspace{0cm} 
\subcaptionbox{Disturbance NN weights\label{fig:dist}}{\includegraphics[width=.32\textwidth,height=9.5cm,keepaspectratio,trim={1.8cm 0.0cm 2cm .08cm},clip]{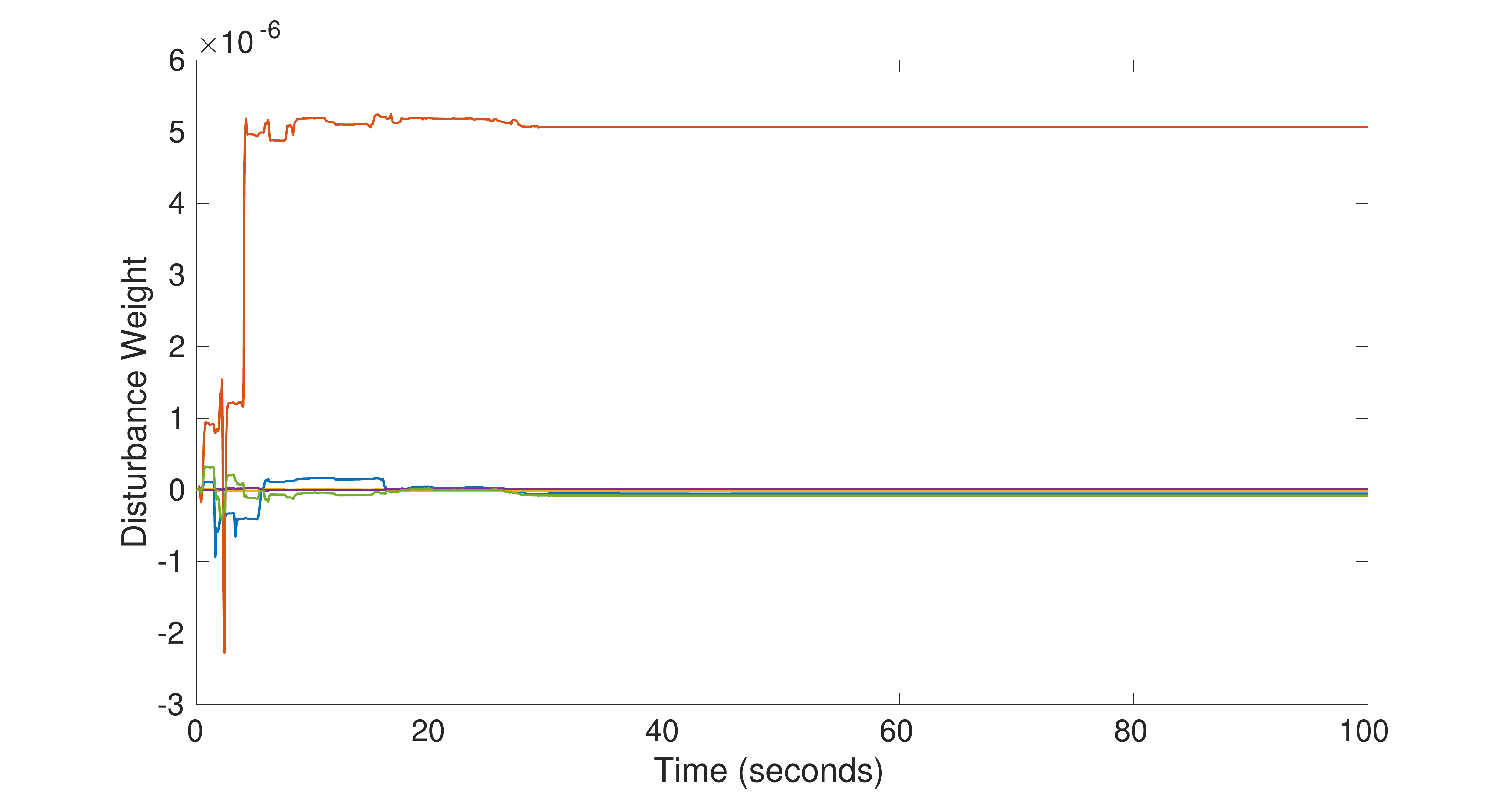}}%
\hspace{0cm} 
\subcaptionbox{Learnt Control Policy\label{fig:ctr}}{\includegraphics[width=.32\textwidth,height=9.5cm,keepaspectratio,trim={1.8cm 0.0cm 2cm .08cm},clip]{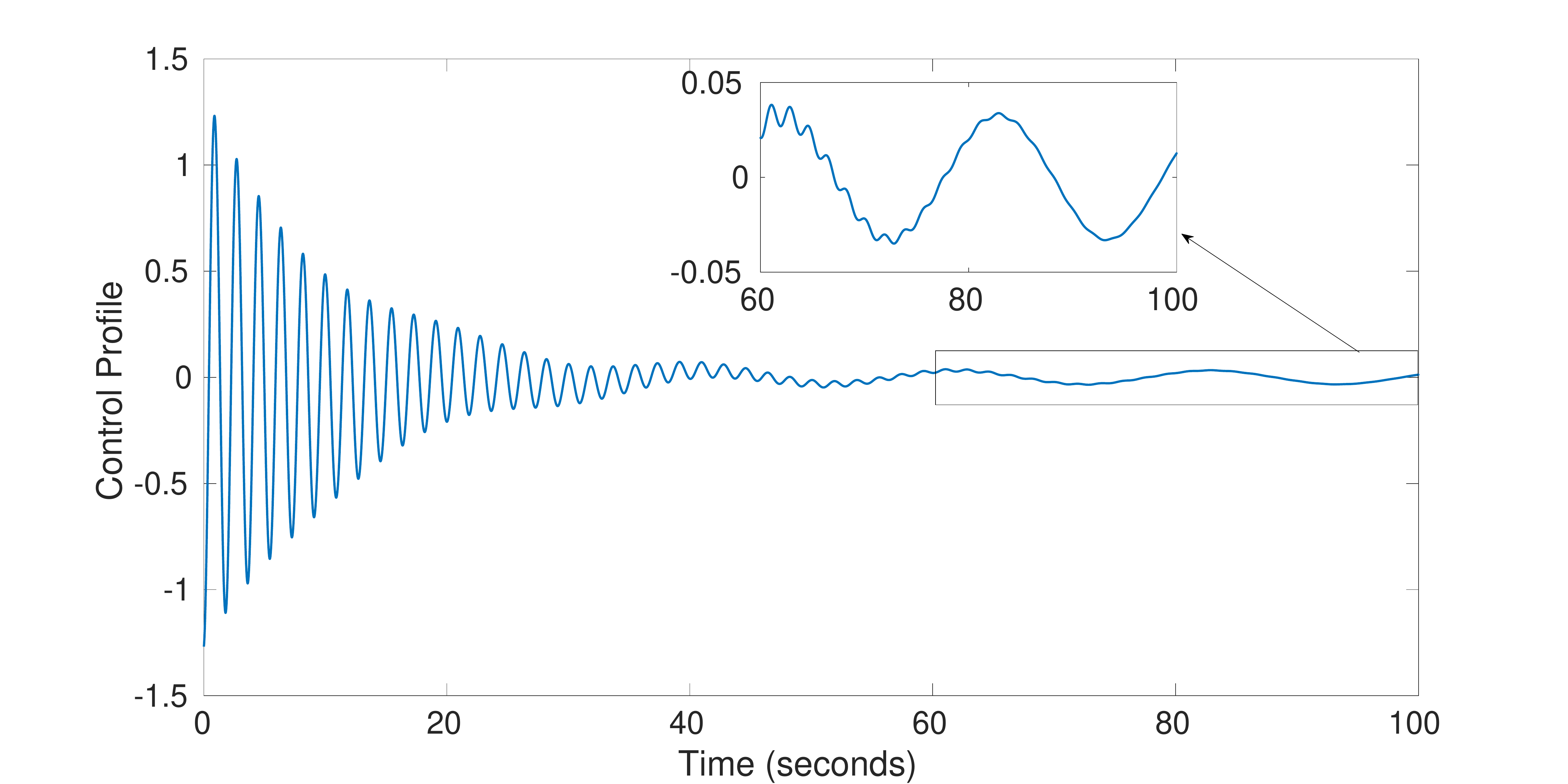}}%
\hspace{0cm}
\subcaptionbox{State Profile under learnt Policy\label{fig:state}}{\includegraphics[width=.32\textwidth,height=9.5cm,keepaspectratio,trim={1.8cm 0.0cm 4cm .08cm},clip]{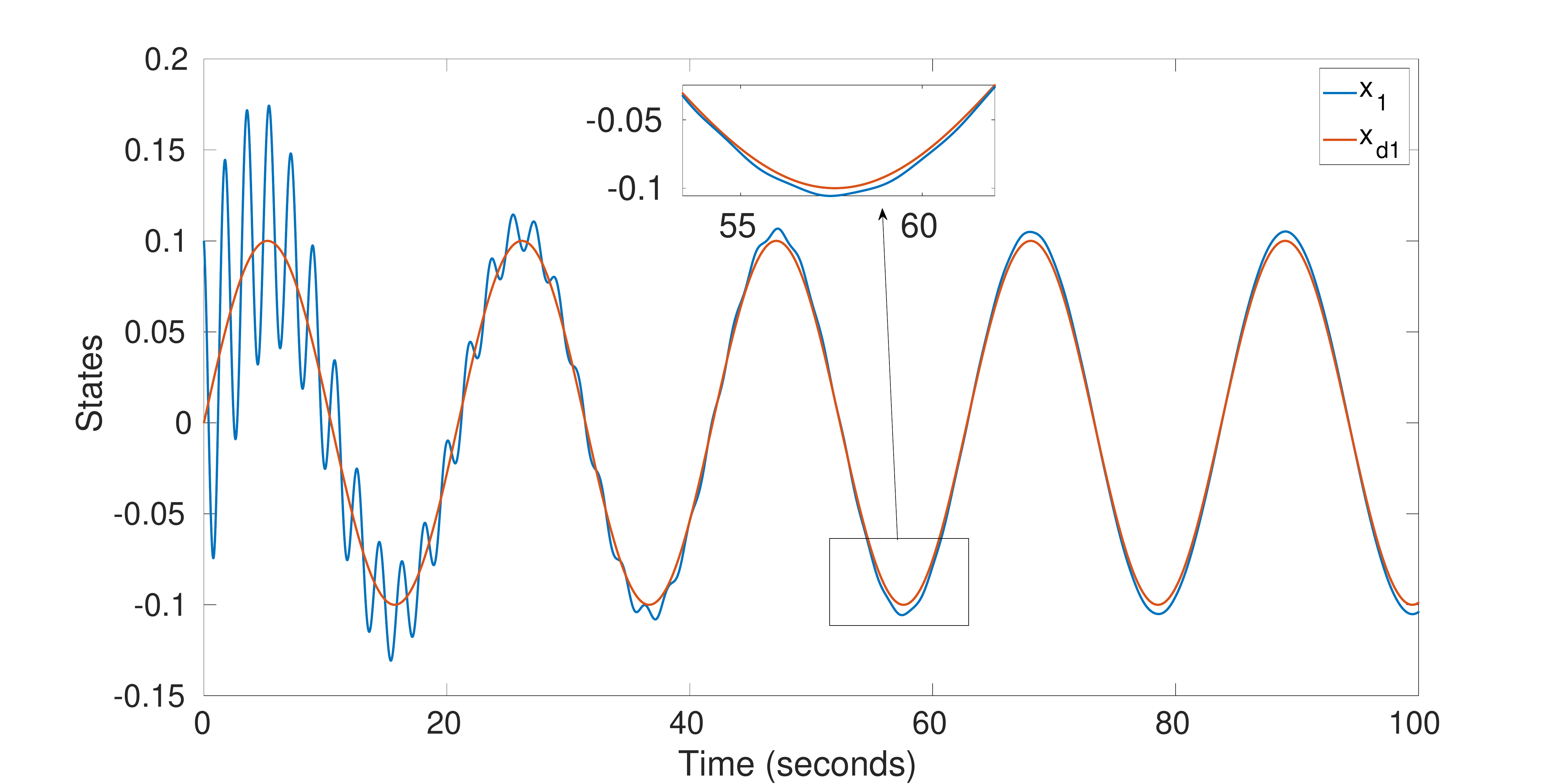}}%
\hspace{0cm} 
\subcaptionbox{HJI error during learning phase\label{fig:err_state}}{\includegraphics[width=.32\textwidth,height=9.5cm,keepaspectratio,trim={1.8cm 0.0cm 2cm .08cm},clip]{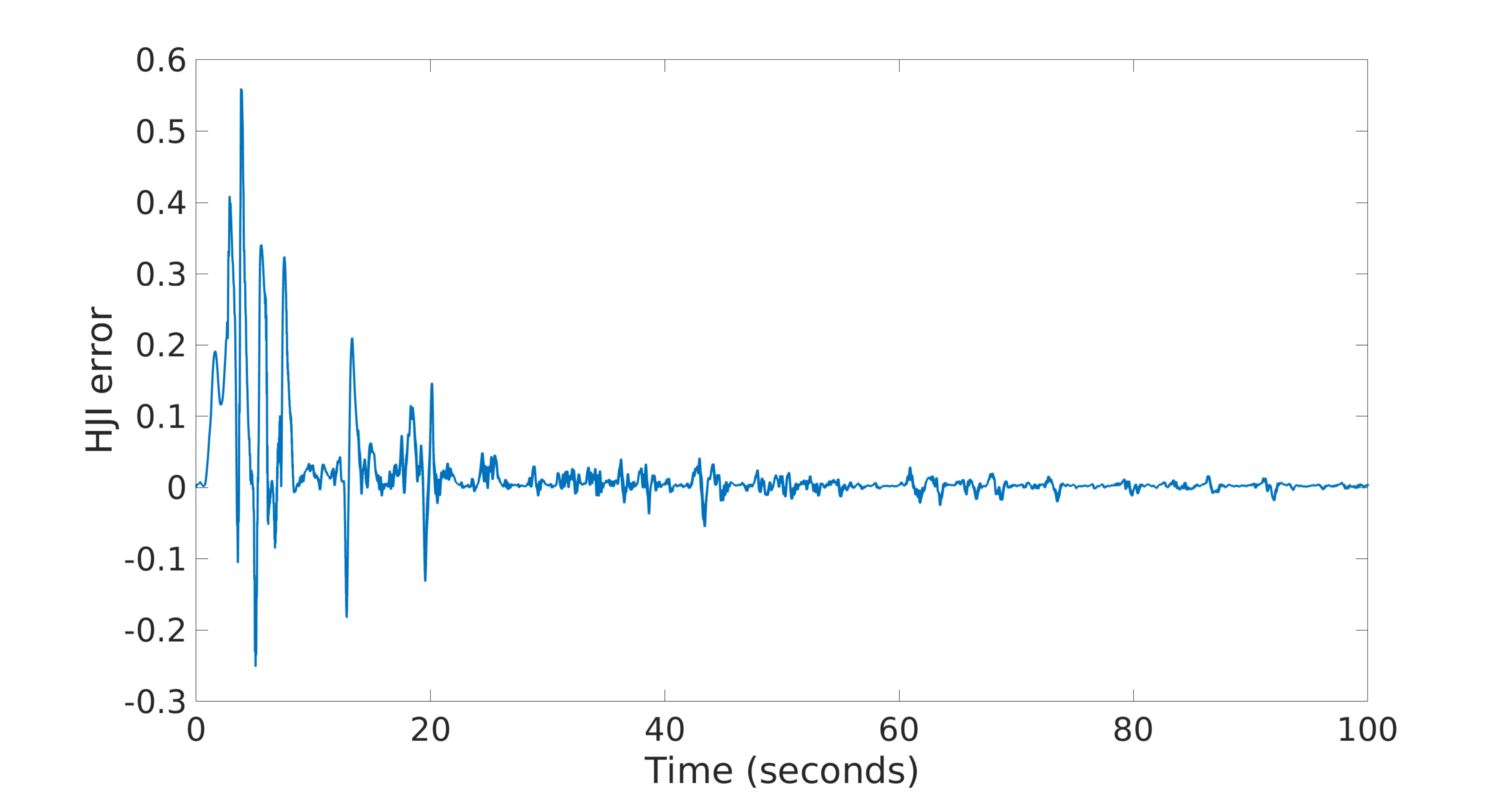}}
\caption{Online training of NN weights and state and control profile without variable gain gradient descent}
\label{fig:NN}
\end{figure*}

\subsubsection{Without variable gain gradient descent scheme}

Simulation results for the update law (\ref{upd_law}) without variable gain gradient descent on the nonlinear system considered above (\ref{nlsys}) are shown in Fig. \ref{fig:NN}. 
For this, the exponent in variable gain term, i.e., $k_1$ was chosen to be $0$, leading to just the constant learning speed (see (\ref{upd_law})) in the online training of the NN weights. 
Note that with $k_1=0$ and $K_1,K_2$ being zero vector and matrix, the update law (\ref{upd_law}) resembles the update law mentioned in \cite{zhu2016using}. 
The difference between (\ref{upd_law}) and the update law of \cite{zhu2016using} arises from the inclusion of disturbance rejection (via the regressor term $\rho$) in (\ref{upd_law}). The NN weights corresponding to critic, actor and disturbance are shown to converge in finite amount of time in Figs. \ref{fig:critic}, \ref{fig:actor}, and \ref{fig:dist}, respectively. 
The final learnt control policy due to the converged weights of critic, actor and disturbance NNs is depicted in Fig. \ref{fig:ctr}.
The tracking error by the learnt policies resulting out of the constant learning rate-based update law is shown to become small in finite time as depicted in Figs. \ref{fig:state} and \ref{fig:err_state}. 
However, It can be observed that there still exists small steady state error in tracking performance (see Fig. 2e). 

\begin{figure*}
\centering
\subcaptionbox{Critic NN weights\label{fig:critic1}}{\includegraphics[width=.32\textwidth,height=9.5cm,keepaspectratio,trim={1.8cm 0.0cm 4cm .08cm},clip]{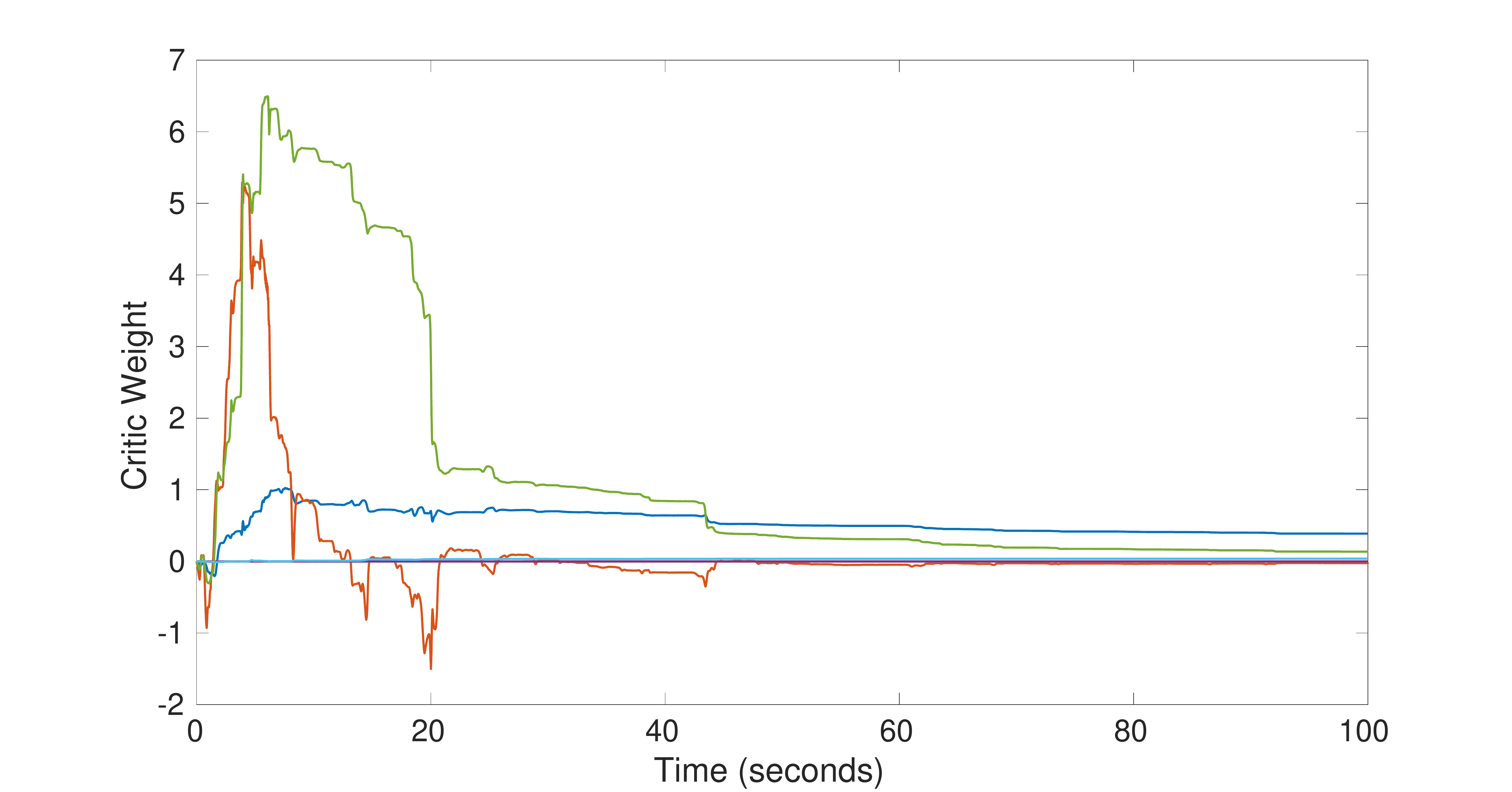}}%
\hspace{0cm} 
\subcaptionbox{Actor NN weights\label{fig:actor1}}{\includegraphics[width=.32\textwidth,height=9.5cm,keepaspectratio,trim={1.8cm 0.0cm 2cm .08cm},clip]{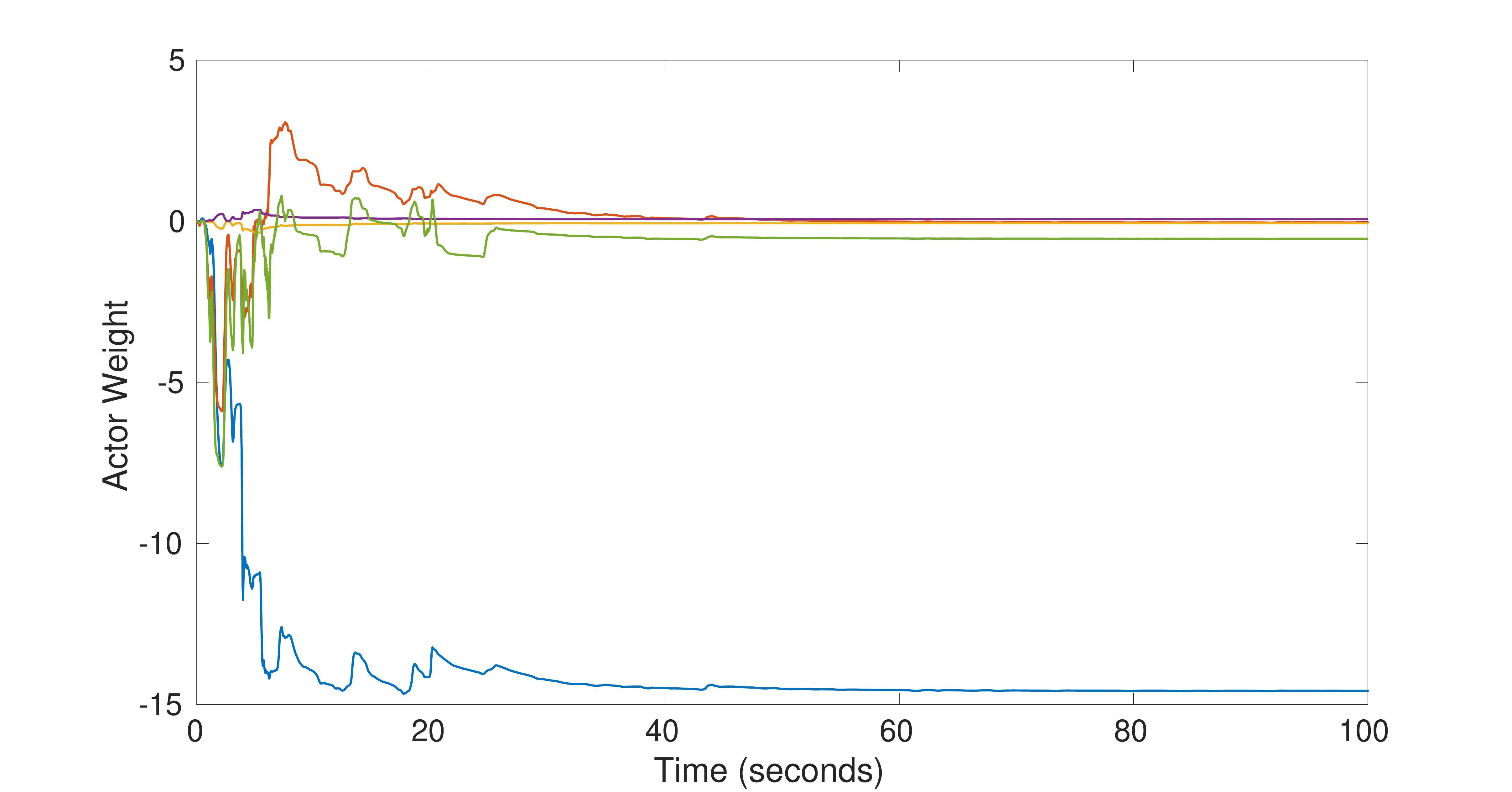}}
\hspace{0cm} 
\subcaptionbox{Disturbance NN weights\label{fig:dist1}}{\includegraphics[width=.32\textwidth,height=9.5cm,keepaspectratio,trim={1.8cm 0.0cm 2cm .08cm},clip]{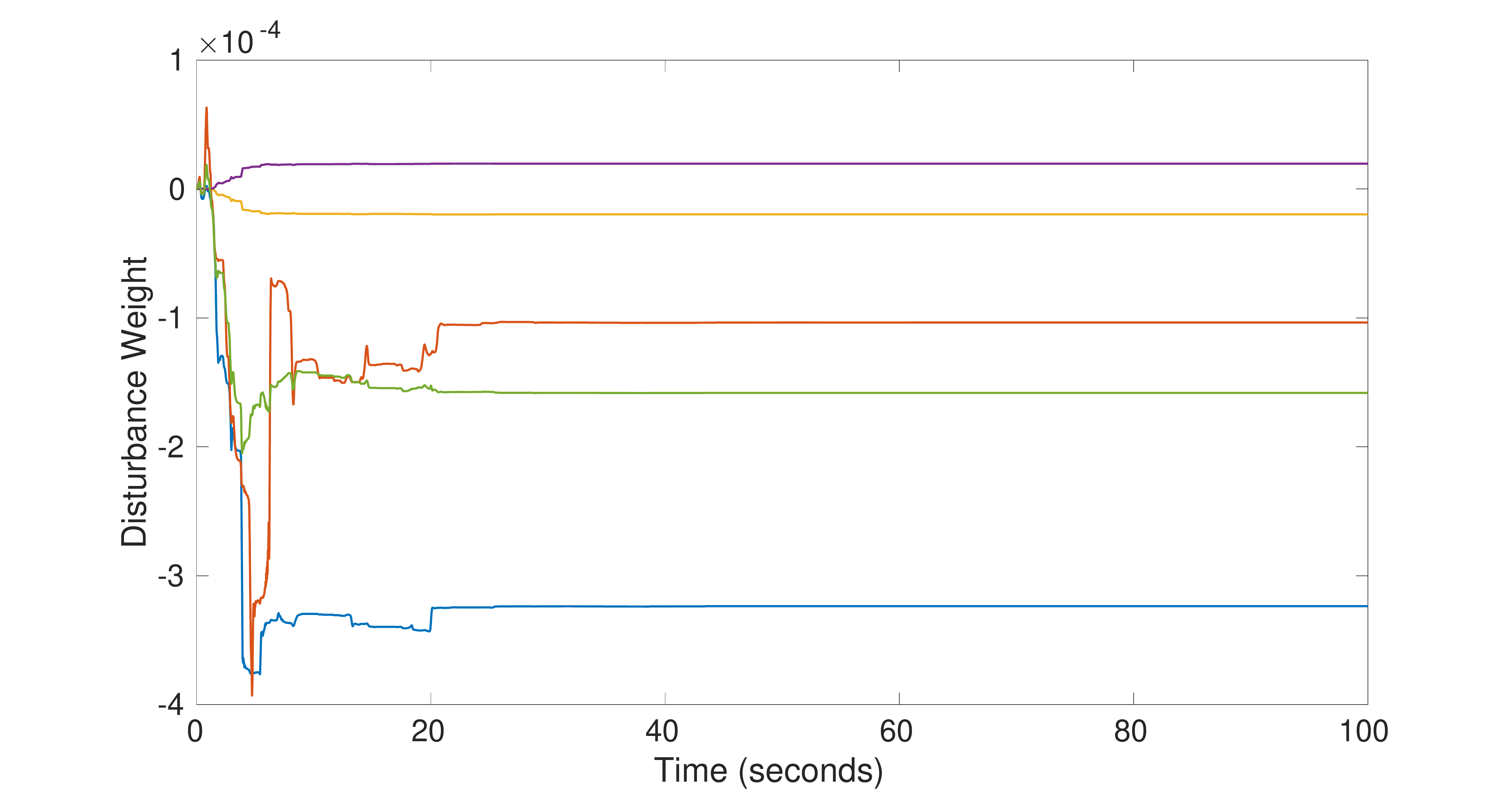}}%
\hspace{0cm} 
\subcaptionbox{Learnt Control Policy\label{fig:ctr1}}{\includegraphics[width=.32\textwidth,height=9.5cm,keepaspectratio,trim={1.8cm 0.0cm 2cm .08cm},clip]{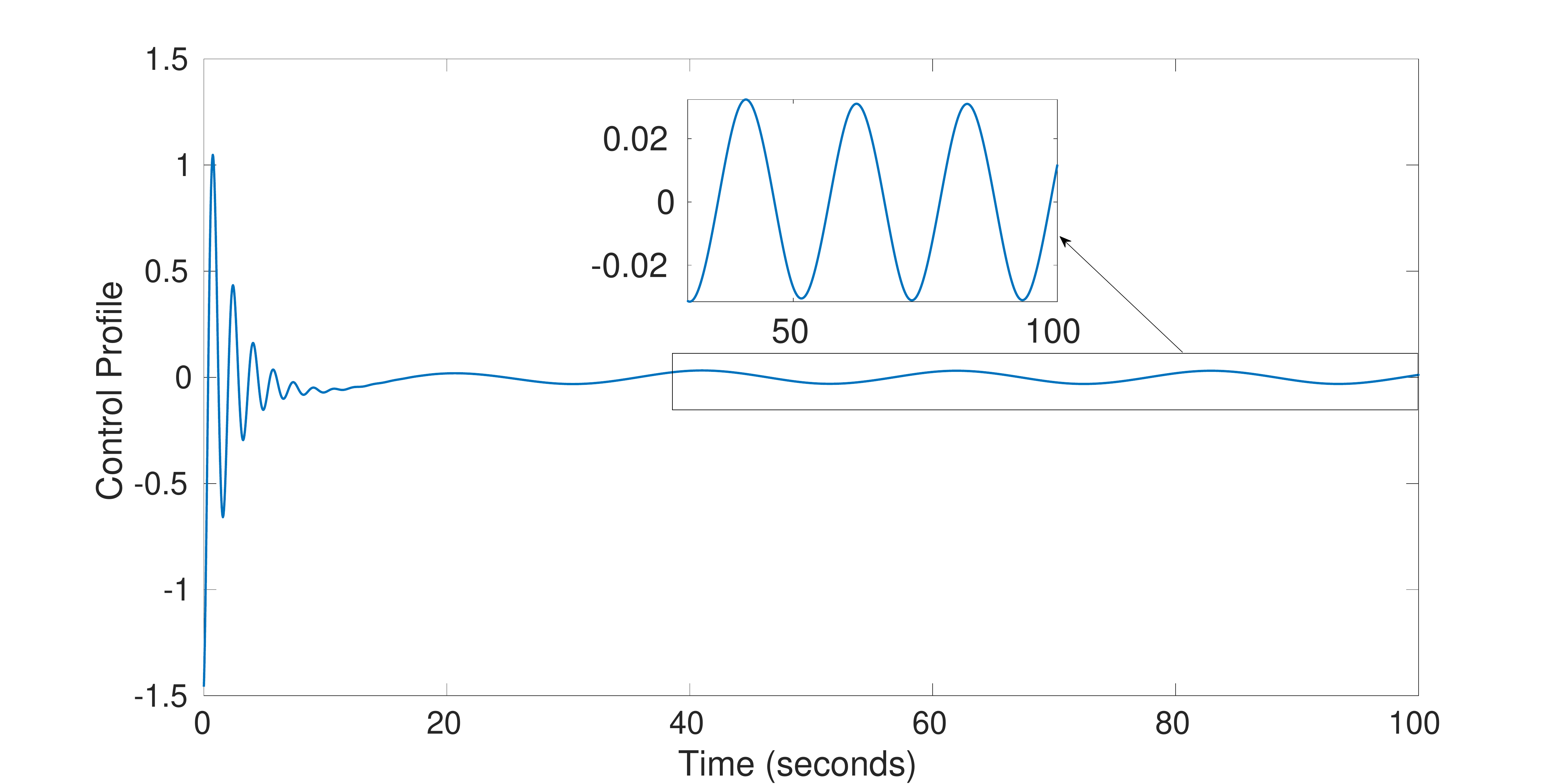}}%
\hspace{0cm}
\subcaptionbox{State Profile under learnt Policy\label{fig:state1}}{\includegraphics[width=.32\textwidth,height=9.5cm,keepaspectratio,trim={1.8cm 0.0cm 4cm .08cm},clip]{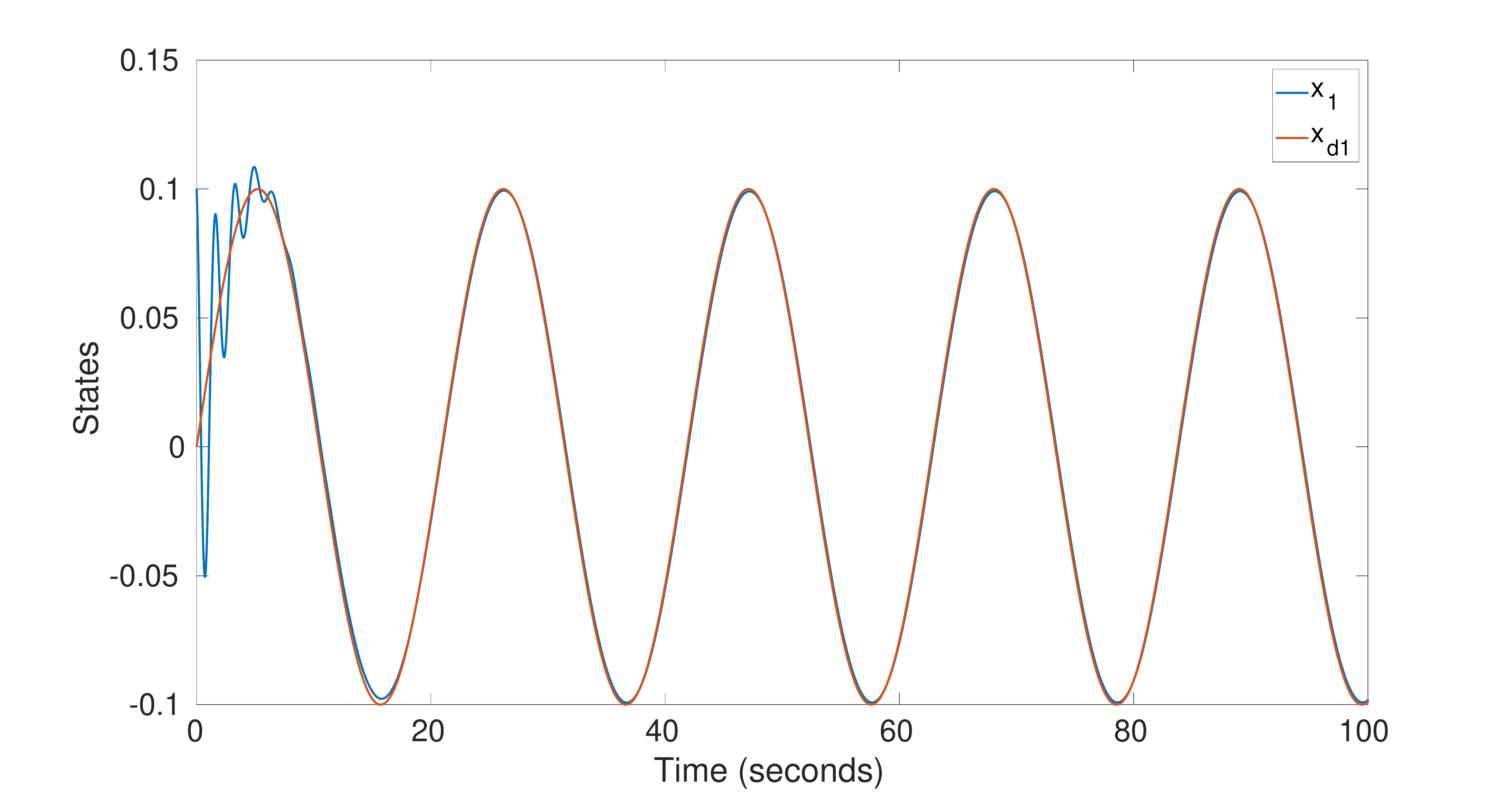}}%
\hspace{0cm} 
\subcaptionbox{HJI error during learning phase\label{fig:err_state1}}{\includegraphics[width=.32\textwidth,height=9.5cm,keepaspectratio,trim={1.8cm 0.0cm 2cm .08cm},clip]{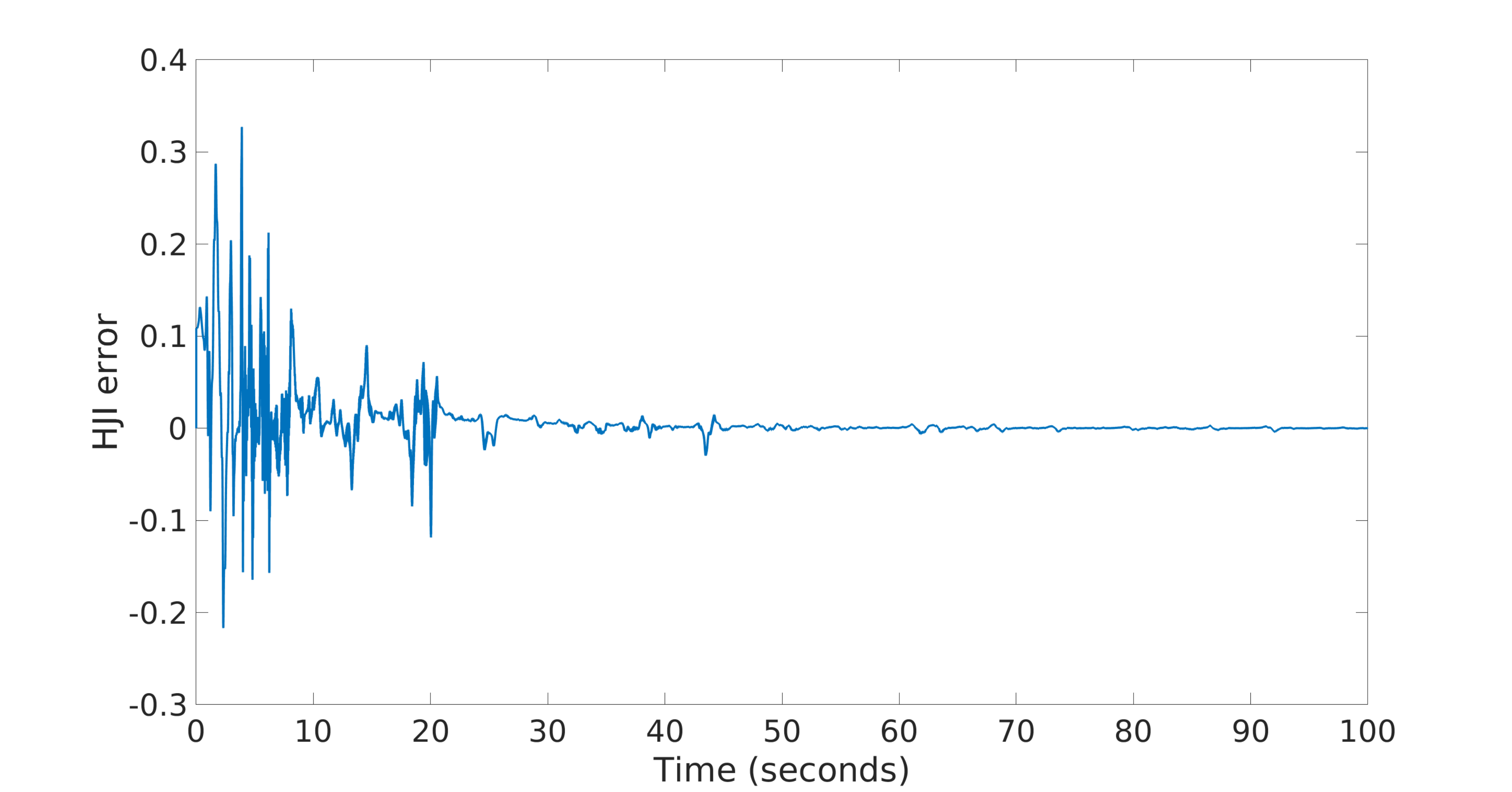}}
\caption{Online training of NN weights and state and control profile with variable gain gradient descent}
\label{fig:sta_ctr1}
\end{figure*}
\subsubsection{With variable gain gradient descent scheme}
Variable gain gradient descent-based update law (\ref{upd_law}) is validated on the nonlinear system (\ref{nlsys}) in Fig. \ref{fig:sta_ctr1}. 
Here, the exponent in variable gain term, i.e., ($k_1$) is chosen to be $0.145$. 
All other parameters are kept same. 
NN weights of critic, actor and disturbance NNs converge very close to their ideal values in a finite amount of time as can be seen in Figs. \ref{fig:critic1}, \ref{fig:actor1} and \ref{fig:dist1}, respectively. 
The learnt control policy arising out of the converged NN weight is depicted in Fig. \ref{fig:ctr1}. 
It is able to track the reference trajectory with high accuracy in finite amount of time as evident from Figs. \ref{fig:state1}. 
The HJI error profile during the learning phase is depicted in Fig. \ref{fig:err_state1} and it can be seen that $|\hat{e}| \leq 1$ during the learning phase.

Note that the oscillations in learnt control policies (see Fig. \ref{fig:ctr}) are more and persist for longer duration in the case when constant learning rate was used as compared to the case when variable gain gradient descent (Fig. \ref{fig:ctr1}) is utilized.
This in turn leads to an oscillatory tracking performance (Fig. \ref{fig:state}) in the transient phase with steady state error for the case with constant learning speed. 
On the other hand, the final learnt policies arising out of variable gain gradient descent-based update law leads to very less oscillations and almost no steady state error (Fig. \ref{fig:state1}).
All this is possible because, the variable gain gradient descent-based update law leads to a much tighter residual set for $\tilde{W}$.
This implies that the control policies resulting out of variable gain gradient descent-based update law are closer to the ideal optimal controller than the policies due to just the constant learning rate gradient descent-based update laws.
It could also be noted from Figs. \ref{fig:err_state} and \ref{fig:err_state1} that the HJI error is within the $[-1,1]$, and since variable gain gradient descent uses a learning rate that is function of instantaneous HJI error, for our problem set, the presence of term $g_1=|\hat{e}|^{k_1}$ actually reduces the learning rate.
This is also the reason why in this case, the convergence time of Fig. \ref{fig:critic1} is slightly longer than Fig. \ref{fig:critic}. 
However, when HJB or HJI error is large, the variable gain gradient descent-based update law leads to faster convergence of NN weights as can be observed in \cite{mishra2019variable}.
\subsection{Linearized F16 Model}\label{lin}
In this section, the tracking performance of the update law developed in this paper (\ref{upd_law}) would be studied on the following linearized dynamics model of F16 fighter aircraft \cite{modares2015h}.
\begin{equation}
\begin{split}
\dot{x}=Ax+Bu+Dd 
\end{split}
\end{equation}
where,
\begin{equation}
\small
 \begin{split}
A=\begin{pmatrix}
-1.01887 & .90506 & -.00215 \\
.82225 & -1.07741 & -.17555 \\
0 & 0 & -1
\end{pmatrix},B=\begin{pmatrix}
0 \\
0 \\
5
\end{pmatrix},D=\begin{pmatrix}
1 \\
0 \\
0
\end{pmatrix}
\end{split}
\end{equation}
The state vector $x=[\alpha,q,\delta_e]^T$, where $\alpha$ is angle of attack,$q$ is the pitch rate and $\delta_e$ is the elevator deflection. The control input is the voltage signal to the elevators and disturbance is caused by the wind gust to the angle of attack.
It is required to track reference angle of attack. For this, the augmented dynamics of $z=[e^T,x_d^T]^T$ is given as,
\begin{equation}
\begin{split}
\dot{z}=A_1z+B_1u+D_1d
\end{split}
\end{equation}
where,  
\begin{equation}
\footnotesize
\begin{split}
A_1=\begin{pmatrix}
A \\
 0_{3\times3}
\end{pmatrix},B_1=\begin{pmatrix}
B \\
0_{3\times1}
\end{pmatrix},~D_1=\begin{pmatrix}
D \\
0_{3\times1}
\end{pmatrix}
\end{split}
\end{equation}
The problem set-up in this section is considered in line with that in \cite{modares2015h}, where the same problem was considered using least-square-based update law. 
The disturbance is assumed to have the form, $d=.1e^{-.1t}\sin{(.1t)}$. Reinforcement interval $T=.001s$ and $R=1,Q=diag([9.9,0,0,0,0,0])$. 
Discount factor $\gamma=.25$ was chosen for simulation.
In the simulation, the desired value of output was $\alpha_d=1.5$ for first $30s$ and then was subsequently changed to $\alpha_d=2.2$ thereafter. 
The constant part of the learning rate is selected as $\eta=209.1$ with variable gain exponent $k_1$ in $|\hat{e}|^{k_1}$ as $k_1=.2$.
The regressor vectors for critic, actor and disturbance NNs were chosen to be,
\begin{equation}
\small
\begin{split}
\sigma_c&=\begin{pmatrix}
z_1z_2,z_1z_3,z_1z_4,z_1z_5,z_1z_6,z_2z_3,z_2z_4,z_2z_5,z_2z_6,z_3z_4,...\\
...z_3z_5,z_3z_6,z_4z_5,z_4z_6,z_5z_6
\end{pmatrix}^T\\
\sigma_a&=\begin{pmatrix}
z_1,z_2,z_2,z_4,z_5,z_6
\end{pmatrix}^T\\
\sigma_d&=\begin{pmatrix}
 z_1,z_2,z_2,z_4,z_5,z_6,z_1z_2,z_1z_3,z_1z_4,z_1z_5,z_1z_6
\end{pmatrix}^T
\end{split}
\end{equation}

The exploratory control policy used during the learning phase is given by $n_1(t)=2e^{(-.009t)}(\sin(t)^2\cos(t)+\sin(3t)^4\cos(1.5t)\\
+\sin(9t)^2\cos(8.4t)+\sin(3.9t)\cos(2.9t)\sin(19t)\\
+\sin(11.9t)\cos(5.3t)^2+\sin(12t)\cos(2.5t)^4+\sin(15t)\cos(1.62t)^2)$

\begin{figure}
     \centering
     \begin{subfigure}[b]{0.4\textwidth}
         \centering
         \includegraphics[width=\textwidth,height=9.5cm,keepaspectratio,trim={2.5cm 0.0cm 4cm 1.08cm},clip]{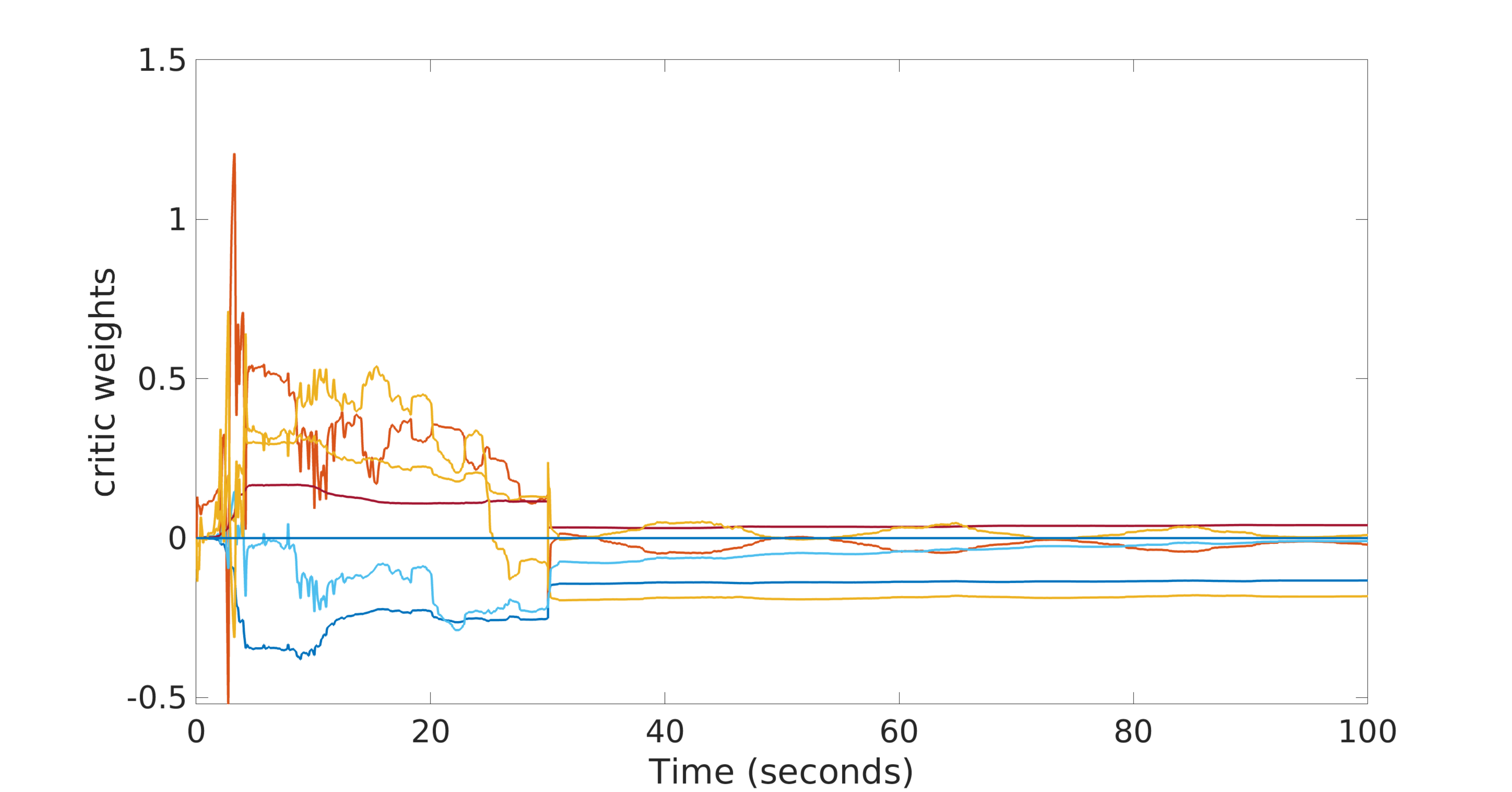}
         \caption{Critic NN Weights for F16 model}
         \label{fig:critic11}
     \end{subfigure}
     \hfill
     \begin{subfigure}[b]{0.4\textwidth}
         \centering
         \includegraphics[width=\textwidth,height=9.5cm,keepaspectratio,trim={2.5cm 0.0cm 4cm 1.08cm},clip]{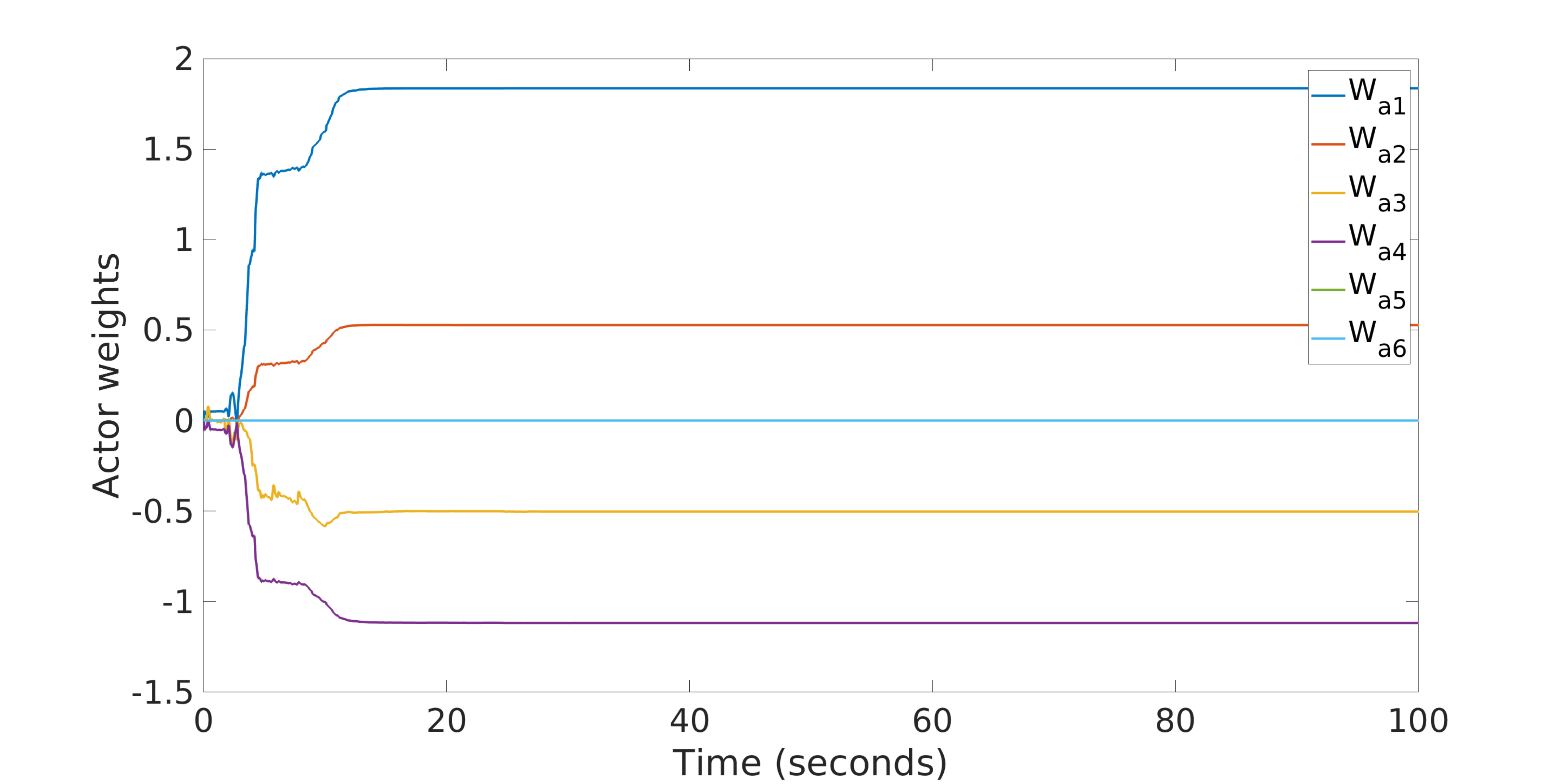}
         \caption{Actor NN Weight for F16 model}
         \label{fig:actor11}
     \end{subfigure}
     \hfill
     \begin{subfigure}[b]{0.4\textwidth}
         \centering
         \includegraphics[width=\textwidth,height=9.5cm,keepaspectratio,trim={2.5cm 0.0cm 4cm .08cm},clip]{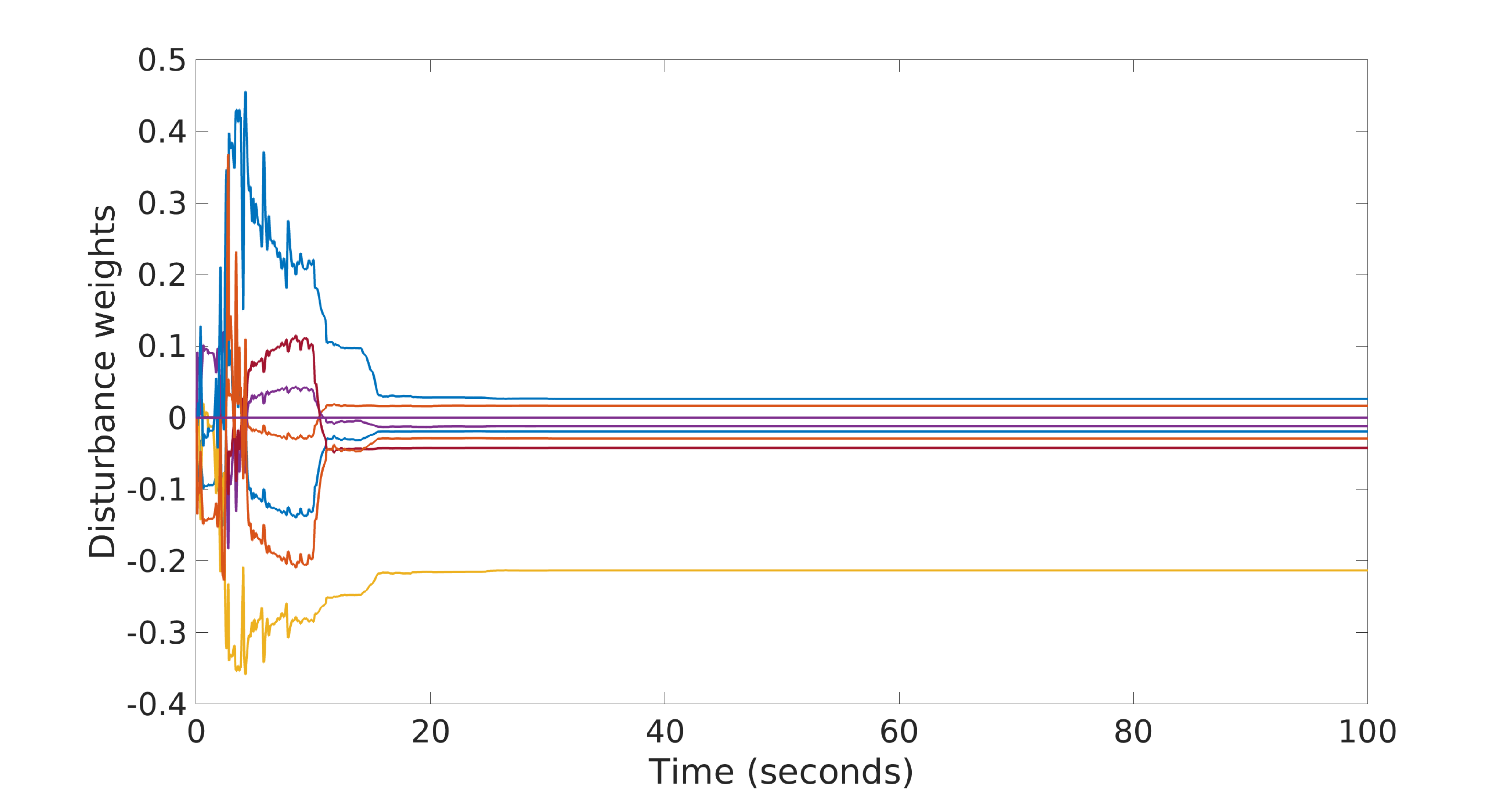}
         \caption{Disturbance NN Weights for F16 model}
         \label{fig:dist11}
     \end{subfigure}
     \hfill
     \begin{subfigure}[b]{0.4\textwidth}
         \centering
         \includegraphics[width=\textwidth,height=9.5cm,keepaspectratio,trim={2.5cm 0.0cm 4cm .08cm},clip]{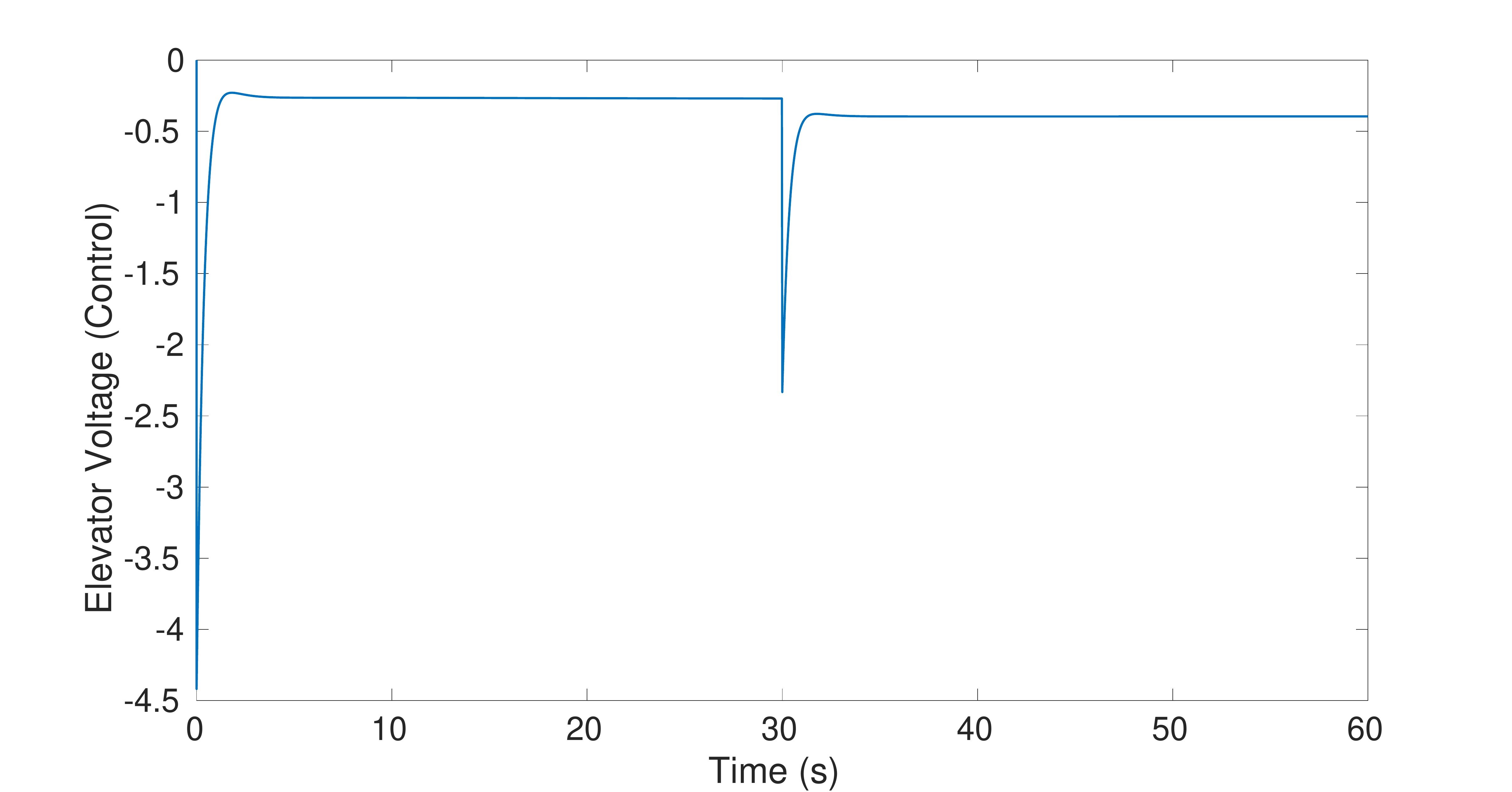}
         \caption{Optimal control policy for F16 model}
         \label{fig:ctr11}
     \end{subfigure}
     \hfill
     \begin{subfigure}[b]{0.4\textwidth}
         \centering
         \includegraphics[width=\textwidth,height=9.5cm,keepaspectratio,trim={2.5cm 0.0cm 4cm .08cm},clip]{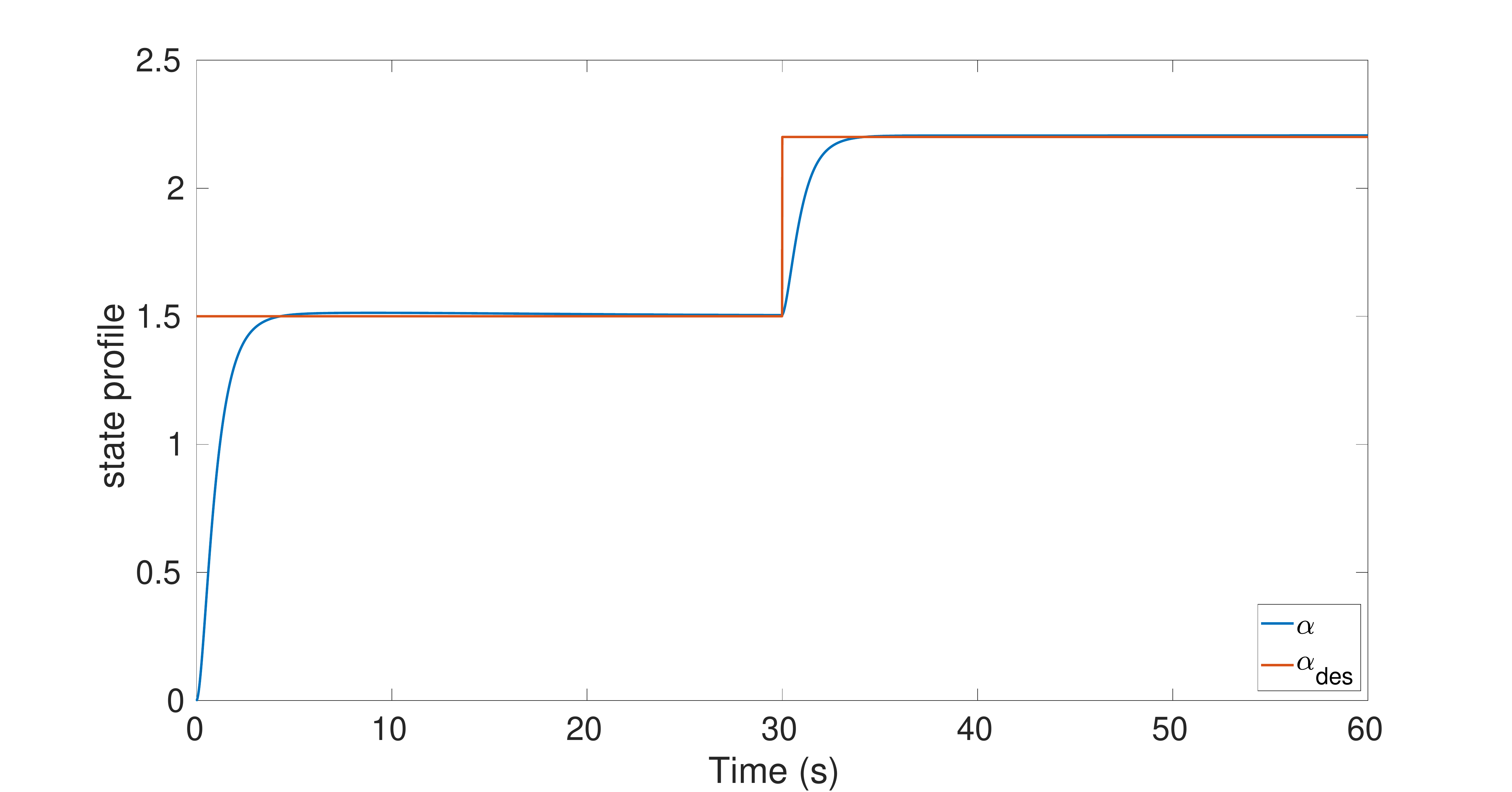}
         \caption{Angle of attack profile under optimal policies}
         \label{fig:alpha1}
     \end{subfigure}
     \vspace{-.2cm}
        \caption{\small Set point tracking with Variable Gain Gradient Descent for F16 model}
        \label{fig:NN11}
\end{figure}

Note from Figs. \ref{fig:critic11}, \ref{fig:actor11}, \ref{fig:dist11} that the NN weights converge close to their ideal values in finite amount of time. 
For this example, in this paper, disturbance attenuation factor is chosen to be $\alpha=1.3$, whereas in \cite{modares2015h} this factor was $\alpha=10$. 
The update law presented in this paper is able to yield optimal policies even under higher degree of disturbance attenuation.
As it can be clearly seen that the final learnt policy (see Fig. \ref{fig:ctr11}) is able to track the set point with high accuracy as is evident from Fig. \ref{fig:alpha1}. 
Contrary to the tracking results presented in \cite{modares2015h} for the same problem set-up, the tracking performance by the presented update law here is devoid of any peak overshoot (refer to Fig. 4 of \cite{modares2015h} and Fig. \ref{fig:alpha1} in this paper). 

Overall, the variable gain gradient descent-based continuous time update law consisting of ER and robust terms lead to much better tracking performance even in the presence of disturbance, which justifies the tighter UUB set proved in Theorem \ref{th1}.

\section{Conclusion}\label{conclusion}
A continuous time neural network (NN) parameter update law driven by variable gain gradient descent, experience replay technique and robust terms for model-free $H_{\infty}$ optimal tracking control problem of continuous time nonlinear system has been presented in this paper. 
Integral reinforcement learning (IRL) has been leveraged in policy iteration framework in this paper. Incorporation of IRL obviates the requirement of drift dynamics in policy evaluation stage, while usage of actor and disturbance NNs to approximate control and disturbance policies obviates the requirement of control coupling dynamics and disturbance dynamics in policy improvement stage. Variable gain gradient descent increases the learning rate when HJI error is large and it dampens the learning rate when HJI error becomes smaller.
It also results in smaller residual set over which the errors in NN weights converge to. Besides this, the ER term and robust terms in the update law help in further shrinking the size of the residual set on which the error in NN weights finally converge to. This results in an improved learnt control policy, sufficiently close to the ideal optimal controller, leading to highly accurate tracking performance.

\section{Appendices}\label{sec14}

\begin{lemma}\label{mm}
Let $x \in \mathbb{R}^n$ and $M \in \mathbb{R}^{n\times n}$ be any square matrix, then, $\lambda_{min}(\frac{M+M^T}{2})\|x\|^2 \leq x^TMx \leq \lambda_{max}(\frac{M+M^T}{2})\|x\|^2$.
\begin{proof}
\begin{equation}
\begin{split}
x^TMx=x^T\Big(\frac{M+M^T}{2}+\frac{M-M^T}{2}\Big)x
\end{split}
\end{equation}
RHS of above equation can be rewritten as:
\begin{equation}
\begin{split}
x^TMx&=x^T(\frac{M+M^T}{2})x+.5x^TMx-.5x^TM^Tx \\
&=x^T(\frac{M+M^T}{2})x+.5x^TMx-.5(x^TMx)^T
\end{split}
\end{equation}
Therefore, 
\begin{equation}
x^TMx=x^T(\frac{M+M^T}{2})x
\label{finalx}
\end{equation}
Using (\ref{finalx}), 
\begin{equation}
\lambda_{min}(\frac{M+M^T}{2})\|x\|^2 \leq x^TMx \leq \lambda_{max}(\frac{M+M^T}{2})\|x\|^2
\end{equation}
\end{proof}
\end{lemma}


\bibliographystyle{iet}
\bibliography{sample}
\end{document}